\documentclass[aps,prx,superscriptaddress,twocolumn]{revtex4-2}

 \usepackage{graphicx,color}
 \usepackage{verbatim}
 \usepackage{amssymb}   
 \usepackage{amsmath}
 \usepackage{amsfonts}
 \usepackage{mathdots}
 \usepackage{hyperref}
 \usepackage{epsfig}
 \usepackage{hyperref}
 \usepackage{xcolor}
 \definecolor{myblue}{RGB}{46, 48,146}
 \hypersetup{colorlinks=true,linkcolor=myblue,citecolor=myblue,urlcolor=myblue, linktocpage}
 \usepackage{braket}
 \usepackage{bm}
 \usepackage{multirow}
 
  \newcommand{\addYM}[1]{#1} 

\begin{document}
 	\title{Theory of Little--Parks oscillations  by vortices in two-dimensional superconductors
 }
 	
	\author{Ying-Ming Xie}\thanks{yingming.xie@sjtu.edu.cn}
    \affiliation{Institute of Condensed Matter Physics, School of Physics and Astronomy, Shanghai Jiao Tong University, Shanghai, China
}
	\affiliation{RIKEN Center for Emergent Matter Science (CEMS), Wako, Saitama 351-0198, Japan} 	
	\author{Naoto Nagaosa} \thanks{nagaosa@riken.jp}
	\affiliation{RIKEN Center for Emergent Matter Science (CEMS), Wako, Saitama 351-0198, Japan} 
	\affiliation{Fundamental Quantum Science Program (FQSP), TRIP Headquarters, RIKEN, Wako 351-0198, Japan} 	

\begin{abstract}

The Little--Parks (LP) effect is a quantum phenomenon in which the superconducting transition temperature of a  superconducting cylinder (or ring) oscillates periodically as a function of the magnetic flux threading the loop.
Recently, multiple experiments have observed half-quantum flux shifts in measurements of LP oscillations, where the oscillations are globally shifted by half a flux quantum compared to conventional cases, a behavior referred to as a $\pi$-ring. Such observations are commonly linked to unconventional pairing symmetries.
In this work, we show that half-quantum flux shifts can arise in two-dimensional (2D) superconducting rings without invoking unconventional pairing symmetry, provided that vortices near the Berezinskii-Kosterlitz-Thouless (BKT) transition are taken into account.
Specifically, based on the vortex-charge duality theory near the BKT transition, we map the problem onto a Coulomb gas model, in which the magnetic flux is represented as a pair of opposite boundary charges (or vortices) at the two edges.
The screening of these boundary charges by thermally excited vortex-antivortex pairs is investigated through explicit Monte Carlo simulations.
Importantly, we demonstrate that the oscillation of the free-vortex density as a function of magnetic flux can exhibit an anomalous half-quantum flux shift.
Our work thus predicts the LP oscillations induced by vortices in 2D superconducting rings near the BKT transition, which provides a new mechanism for generating $\pi$-rings.

 	\end{abstract}
 	
 	\date{\today}
 	
 	\maketitle

\emph{Introduction.---} 
Superconducting quantum phenomena play important roles in quantum computation and quantum information processing.
One of the earliest experiments demonstrating the quantum nature of superconductors is the observation of quantized magnetic flux trapped in units of the flux quantum $\Phi_0 = h/(2e)$ in a multiply connected superconductor \cite{Deaver1961, Doll1961}, where $h$ is Planck’s constant and $e$ is the elementary charge.
Following these experiments, Little and Parks studied thin-walled superconducting cylinders subjected to a magnetic field flux and showed that the resistance oscillates with a period of $\Phi_0$, due to the periodic change in superconducting transition temperature $T_c$ \cite{Little1962, Park1964}.
This Little--Parks (LP) effect can be understood within Ginzburg--Landau theory \cite{tinkham2004, grosso2013solid}, in which the magnetic field enters as a gauge potential, analogous to the Aharonov--Bohm effect.


The LP effect has also been  used as a probe of unconventional superconductivity.
Following the discovery of high-temperature superconductors and heavy-fermion superconductors, a half-quantum flux shift, or equivalently a $\pi$-phase shift, in a superconducting ring was proposed as a diagnostic of spin-triplet pairing or other unconventional pairing symmetries \cite{Tsuei2000, Manfred2005, GLB1987}.
Especially,  chiral spin-triplet pairing can generate an intrinsic and topologically protected phase winding in the superconducting condensate \cite{Kim2007, Victor2009}, which can lead to a $\pi$-phase shift in the LP effect analogous to the half-quantum vortices observed in the $^3$He-A phase \cite{Volovik1985}.
Recently, multiple experiments have reported $\pi$-phase shifts or half-quantum flux states in superconducting rings without junction regions, including $\beta$-Bi$_2$Pd \cite{Yufan2019}, $\alpha$-BiPd \cite{Xiaoying2020}, TaS$_2$ \cite{Wan2024, Almoalem2024}, Bi/Ni bilayers \cite{Tokuda2025}, CsV$_3$Sb$_5$ \cite{benchuan2025}, and 2M-WS$_2$ \cite{ZhangTRSB}.
Notably, these experiments commonly suggest that the emergence of half-quantum flux shift is associated with the possible presence of chiral or spin-triplet superconducting pairing.
These experimental advances have also stimulated renewed theoretical interest in this phenomenon \cite{Aoyama2022, Lee2023, Hua2023, LiYi2025}.

However, an important aspect has so far been overlooked.
These recent experiments are typically performed on thin flakes, being close to the two-dimensional (2D) limit. In 2D superconductors, the resistance oscillations are measured at temperatures near the onset of finite resistance, which is usually close to the BKT  transition temperature $T_{\mathrm{BKT}}$.
This raises a crucial question: how does the LP effect behave in the presence of vortices near the BKT transition in 2D superconductors?
In particular, can the anomalous half-quantum flux shift be driven by BKT physics without involving unconventional pairing?

In this work, we present a systematic study of the LP effect in a conventional 2D superconducting cylinder or ring near the BKT transition (the geometry is schematically illustrated in Fig.~\ref{fig:fig1}(a) and (b)).
We first construct a vortex--charge duality theory and map the problem onto a 2D neutral Coulomb gas model, in which the magnetic flux is mapped to boundary charges [Fig.~\ref{fig:fig1}(c)].
The bulk charges tend to screen these boundary charges in order to reduce the effective electric field in the bulk.
We explicitly perform Monte Carlo simulations to investigate the effects of the boundary charges.
Near the BKT transition, resistance oscillations are expected to be proportional to the density of free vortices.
As the magnetic flux modifies the boundary charges with a period of $\Phi_0$, we find that the free-vortex density exhibits periodic oscillations, leading to the LP behavior in practice.
Importantly, we demonstrate how the half-quantum flux shift can appear in 2D superconducting cylinder/rings  over a wide parameter regime through field-modified vortex dynamics near the BKT transition [see illustrations in Fig.~\ref{fig:fig1}(c) and (d)].
In this mechanism, the boundary charges act as effective pinning sources that reduce the density of free vortices, resulting in resistance peaks at integer flux.
Our theory therefore establishes a  BKT mechanism for half-quantum flux shift in superconducting rings and advances the understanding of the LP effect in 2D superconductors.

\begin{figure}
    \centering
    \includegraphics[width=1\linewidth]{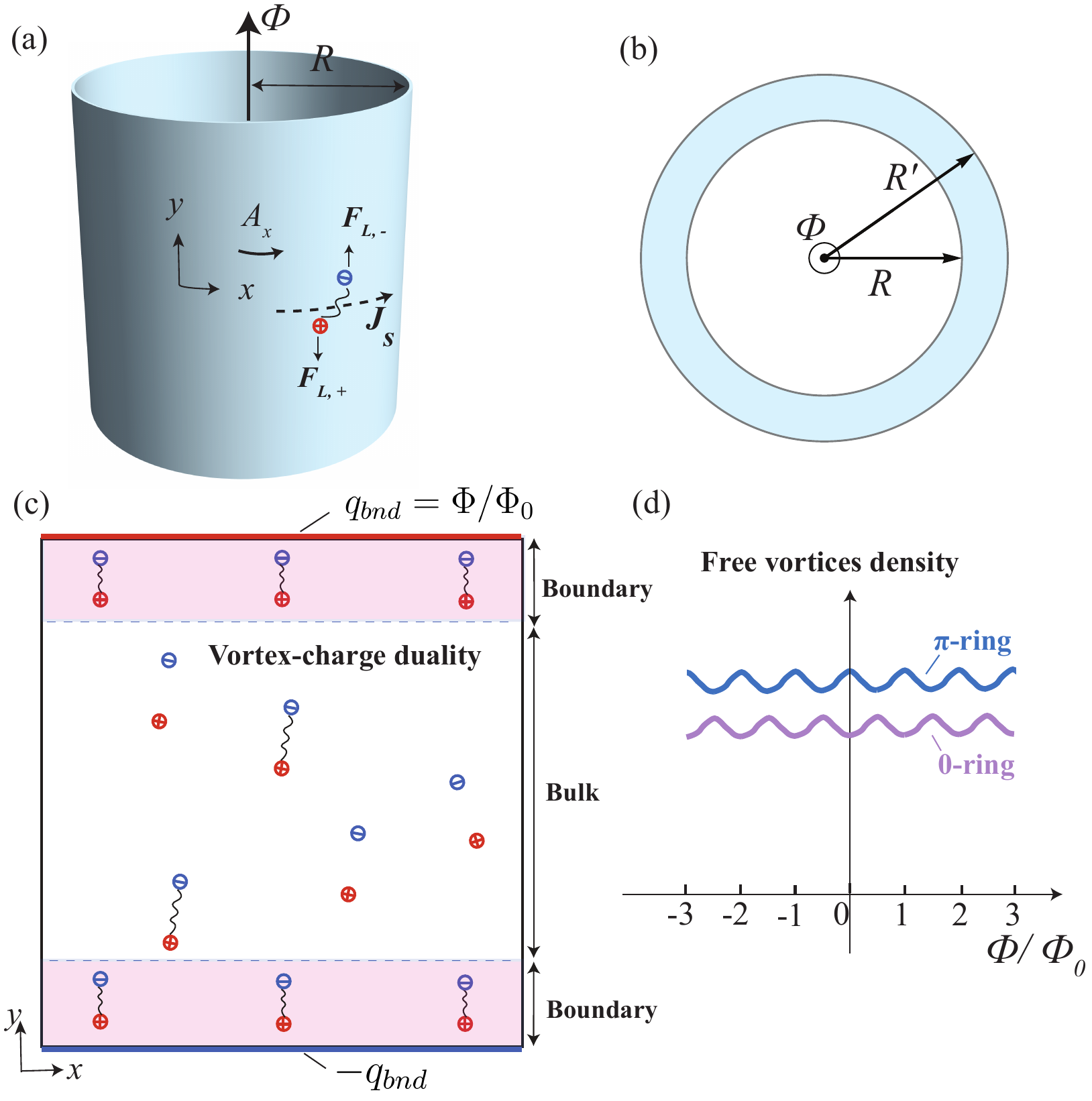}
    \caption{
    (a,b) Schematic illustrations of a hollow superconducting cylinder and a ring, where the wall thickness is in the 2D limit. A magnetic flux $\Phi$ threads the central hole of the cylinder or ring, with radii labeled $R$ and $R'$, respectively.  The $\bm{F}_{L,\pm}$ labels the Lorentz forces. 
(c) Representation of the cylinder in panel (a) after the vortex--charge duality mapping, where the $x$ direction is periodic and the $y$ direction is open. The total boundary charge $q_{bnd}= \pm \Phi/\Phi_0$ uniformly spreads at top and bottom, respectively (red and blue indicate opposite charges). Note that a shift by $n\Phi_0$ can be absorbed into the phase winding around the hole of the cylinder. Charge pairs represent vortex–antivortex pairs, with bound pairs indicated by those connected by a curved line. Near the boundaries (pink), vortex charges tend to form aligned dipoles that screen the boundary charges.
(d) Free-vortex density as a function of magnetic flux at $0$- and $\pi$-ring cases.
}
    \label{fig:fig1}
\end{figure}

\emph{Duality theory for 2D superconductors near the BKT transition.---} For completeness,  we begin by presenting the field theory near the BKT transition in the presence of external gauge fields. We start from a ((2+1))D action and map the theory into a 2D classical Coulomb gas, making its generalization to ring or cylindrical geometries more transparent. 
The partition function is given by $Z=\int D[\theta(r,\tau)] \exp[-S]$, and the action $S$ is  \cite{nagaosa2013quantum}
\begin{eqnarray}
S&&=\int_{0}^{\beta}d\tau \int d^2\bm{r} \frac{J}{2}[(\partial_{\tau}\theta)^2+(\nabla \theta)^2].
\end{eqnarray}
Here $\tau$ is the imaginary time, $\beta=1/(k_B T)$, $\bm{r}=(x,y)$, and $J$ is the phase stiffness.
The action arises from phase fluctuations of the superconducting order parameter,
\begin{equation}
\Delta(\bm{r},\tau)=|\Delta(\bm{r},\tau)|e^{i\theta(\bm{r},\tau)}=\Delta_0 e^{i\theta(\bm{r},\tau)}.
\end{equation}
The phase field $\theta(\tilde{x})$, with $\tilde{x}=(\bm{r},\tau)$, can be separated into a smooth part $\theta_0(\tilde{x})$ and a singular vortex contribution $\theta_v(\tilde{x})$ \cite{nagaosa2013quantum},
\begin{equation}
\theta(\tilde{x})=\theta_0(\tilde{x})+\theta_v(\tilde{x}).
\end{equation}
For example, for a vortex core located at $\bm{X}=(X_1,X_2)$, the singular part is
$\theta_v(\bm{r})=\arctan[(r_2-X_2)/(r_1-X_1)]$.
The vortex current is given by
$j_{\mu}=\frac{1}{2\pi}\epsilon_{\mu\nu\lambda}\partial_{\nu}\partial_{\lambda}\theta_{v}(\tilde{x})$,
which satisfies
$(\partial_1\partial_2-\partial_2\partial_1)\theta_v(\bm{r})=2\pi \delta(\bm{r}-\bm{X}(\tau))$ \cite{nagaosa2013quantum}.
Here the indices $\mu,\nu,\lambda=0,1,2$ label $(\tau,x,y)$.
In particular, $j_0(\bm{r},\tau)=\delta(\bm{r}-\bm{X}(\tau))$ gives the vortex density, while
$j_{\alpha}(\bm{r},\tau)=\partial_{\tau} X_{\alpha}(\tau)\delta(\bm{r}-\bm{X}(\tau))$ describes the spatial vortex current.

Next, we introduce the gauge potential into the action and establish the dual theory via a Hubbard--Stratonovich (HS) transformation.
Specifically, the action becomes
\begin{equation}
    S=\int d^3\tilde{x} \frac{J}{2}(\partial_{\mu}\theta+2eA_{\mu})^2,
\end{equation}
where $A_{\mu}$ denotes the gauge potential.
After the HS transformation, the partition function takes the form
$Z=\int D\theta_0 D\theta_{v} DJ_{\mu}\exp[-\tilde{S}]$, with
\begin{equation}
  \tilde{S}=\int d^3\tilde{x}(\frac{J_{\mu}^2}{2J}+iJ_{\mu}[\partial_{\mu}\theta_0+\partial_{\mu}\theta_{v}+2eA_{\mu}]).
\end{equation}
Integrating out $\theta_0$ imposes the conservation law $\partial_{\mu} J_{\mu}=0$, which can be solved as
$J_{\mu}= \frac{1}{2\pi}\epsilon_{\mu\nu\lambda}\partial_{\nu}a_{\lambda}$.
For compactness, we denote $J_{3D}=\frac{1}{2\pi} \nabla_{3D}\times a_{3D}$.
It is straightforward to show that
$\int d^3\tilde{x}  J_{\mu}\partial_{\mu}\theta_{v}=\int d^3\tilde{x}a_{\lambda}j_{\lambda}$ and
$2e \int d^3\tilde{x}J_{\mu} A_{\mu}=\frac{e}{\pi}\int d^3\tilde{x}  a_{\lambda} B_{\lambda}$.
The resulting total action is
\begin{eqnarray}
\tilde{S'}&&= \mathcal{A}\sum_{i}\int  d s_i+ \int d^3\tilde{x} \frac{(\nabla_{3D}\times a_{3D})^2}{8\pi^2J}\nonumber\\
&&+i\int d^3\tilde{x}  a_{3D}\cdot (j_{3D}+\frac{e}{\pi}B_{3D}). \label{total_ac}
\end{eqnarray}
The first term on the right-hand side represents the energy cost of vortex cores, with action $\mathcal{A}$ per unit length multiplied by the length of the world line $\int d s_i$ of the $i$th vortex.
In the static 2D limit, $a_{3D} \mapsto a_0(\bm{r})$, $j_{\mu}\mapsto j_0=\rho_v(\bm{r})$, and $B_{3D}\mapsto B_0=B_z$, yielding
\begin{equation}
\tilde{S}'=\beta \int d^2 \bm{r} \{ \mathcal{A}|\rho_{v}(\bm{r})|+i a_0(\bm{r})(\rho_v(\bm{r})+\frac{e}{\pi} B_z)+ \frac{|\nabla a_0(\bm{r})|^2}{8\pi^2J}\}. \label{dual_action}
\end{equation}

The action $\tilde{S'}$ describes a 2D neutral Coulomb gas, where $\rho_v(\bm{r})$ denotes the vortex charge density.
As an illustration, we set $B_z=0$ and write the vortex density as
$\rho_{v}=\sum_{i} q_i\delta(\bm{r}-\bm{X}_i)$,
where $i$ labels the $i$th vortex and $q_i=\pm 1$ corresponds to a vortex or antivortex.
By integrating out the scalar field, an effective action is obtained [see Supplementary Material (SM) \cite{Supp}]
\begin{eqnarray}
\tilde{S}_{eff}=\frac{1}{2}\sum_{i\neq j} q_i q_jV(\bm{X}_{ij})+ E_c\sum_{i}q_i^2, \label{gas}
\end{eqnarray}
where $\bm{X}_{ij}=\bm{X}_i-\bm{X}_j$, $E_c$ is the vortex core energy, and the interaction takes the 2D logarithmic Coulomb form
$V(\bm{X}_{ij})=-2\pi J \ln \frac{|\bm{X}_{ij}|}{r_c}$, with $r_c$ a short-distance cutoff.
Defining the charge unit as $\sqrt{2\pi J}$, $\tilde{S}_{eff}$ is explicitly identified as a 2D neutral Coulomb gas model.

\emph{Dual-field theory description of the LP effect for 2D superconductors.---} 
In the LP effect, a superconductor forms a ring of radius $R$ threaded by a magnetic flux $\Phi$.
As shown in Fig.~\ref{fig:fig1}(a), we consider a 2D superconductor with cylindrical topology, where the $x$ direction is periodic and the $y$ direction is open.
Inside the sample, $B_z=0$, while the vector potential $\bm{A}=\frac{\Phi}{2\pi R}\hat{e}_x$ remains finite, analogous to the Aharonov--Bohm effect.
A narrow 2D ring ($|R'-R|/R\ll1$), illustrated in Fig.~\ref{fig:fig1}(b), is geometrically equivalent to the system in Fig.~\ref{fig:fig1}(a).
This equivalence is obtained by replacing the system length along the $y$ direction, $L_y$, in Fig.~\ref{fig:fig1}(a) with the ring width $|R'-R|$ in Fig.~\ref{fig:fig1}(b).
The central question is how such a vector potential modifies the duality theory and whether it leads to novel physical consequences for the LP effect in 2D superconductors near the BKT transition.

Next, we show that the vector potential in the ring or cylinder geometry of 2D superconductors induces boundary vortex charges in the dual theory.
Specifically, starting from the dual action, the coupling to the external field in the case of $\nabla_{3D}\times A_{3D}=0$ can be reduced to a boundary term [see Sec. I(C) of the SM for a detailed derivation],
\begin{equation}
   \frac{ie}{\pi} \int d^3\tilde{x} (\nabla_{3D}\times a_{3D})\cdot A_{3D}=\frac{ie \beta}{\pi} \int _{\partial V} d^2\bm{r}\cdot ( a_{3D} \times A_{3D}), \label{Eq_8}
\end{equation}
where we have used the vector identity
$\text{div}(a_{3D} \times A_{3D})=A_{3D}\cdot (\nabla_{3D} \times a_{3D})-a_{3D}\cdot (\nabla_{3D} \times A_{3D})$.
The vector product is given by
$a_{3D} \times A_{3D}= (a_x A_y- a_y A_x, a_y A_0- a_0 A_y, a_0 A_x- a_x A_0)$. In the cylindrical geometry, $A_0=0$ and $(A_x, A_y)=(\frac{\Phi}{2\pi R},0)$.
We find that the right-hand side of Eq.~\eqref{Eq_8} reduces to a boundary term
$i\beta \int d^2\bm{r} \rho_{v}^{(A)} a_0(\bm{r})$, with the boundary charge
\begin{equation}
\rho_v^{(A)}(\bm{r})=\frac{q_{bnd}}{L_x}[\delta(y-L_y)-\delta(y)]. \label{Eq_bc}
\end{equation}
Here, the boundary charge $q_{\mathrm{bnd}}=\frac{\Phi}{\Phi_0}$,
$\Phi_0=\frac{\pi}{e}$ is the superconducting flux quantum (with $\hbar=1$), and the cylinder perimeter is $L_x=2\pi R$.
The term $\rho_v^{(A)}(\bm{r})$ represents boundary charges $-q_{\mathrm{bnd}}$ and $q_{\mathrm{bnd}}$ uniformly distributed along the edges at $y=0$ and $y=L_y$, respectively [Fig.~\ref{fig:fig1}(c)].
Since a superconducting ring threaded by $\Phi=n\Phi_0$ can be absorbed into as a  phase winding around the hole of the cylinder, we redefine the boundary charge as
$q_{\mathrm{bnd}}=\frac{\Phi}{\Phi_0}-n$, where $n$ is chosen such that
$\left|\frac{\Phi}{\Phi_0}-n\right|<\frac{1}{2}$.

Remarkably, when $\Phi \neq n\Phi_0$, the boundary charge $q_{\mathrm{bnd}}$ is finite and fractional.
The opposite boundary charge at two edges generates electric fields in the bulk, which in turn induce net positive and negative charge accumulation near the two boundaries, as we would show later.
The corresponding dual picture  is that the magnetic flux generates a supercurrent that produces Lorentz forces, thereby driving vortex dynamics, as illustrated in Fig.~\ref{fig:fig1}(a). In other words, this boundary charge represents the Lorentz force acting on the vortices due to the superconducting current induced by the flux.

By replacing the field-dependent term in Eq.~\eqref{dual_action} with the boundary charge term, we find that the dual description of the LP effect near the BKT transition is given by
\begin{equation}
\tilde{S}_{LP}=\beta \int d^2 \bm{r} \{ \mathcal{A}|\rho_{v}|+i a_0(\rho_v+\rho_{v}^{(A)})+ \frac{|\nabla a_0|^2}{8\pi^2J}\}.
\end{equation}
The action $\tilde{S}_{LP}$ describes a Coulomb gas problem with an additional boundary charge $\rho_{v}^{(A)}$. Compared with previous formulations \cite{nagaosa2013quantum, Fisher}, the present theory explicitly incorporates the external flux in a cylinder/ring geometry, whose effects are the focus of this work.

\begin{figure}
    \centering
    \includegraphics[width=1\linewidth]{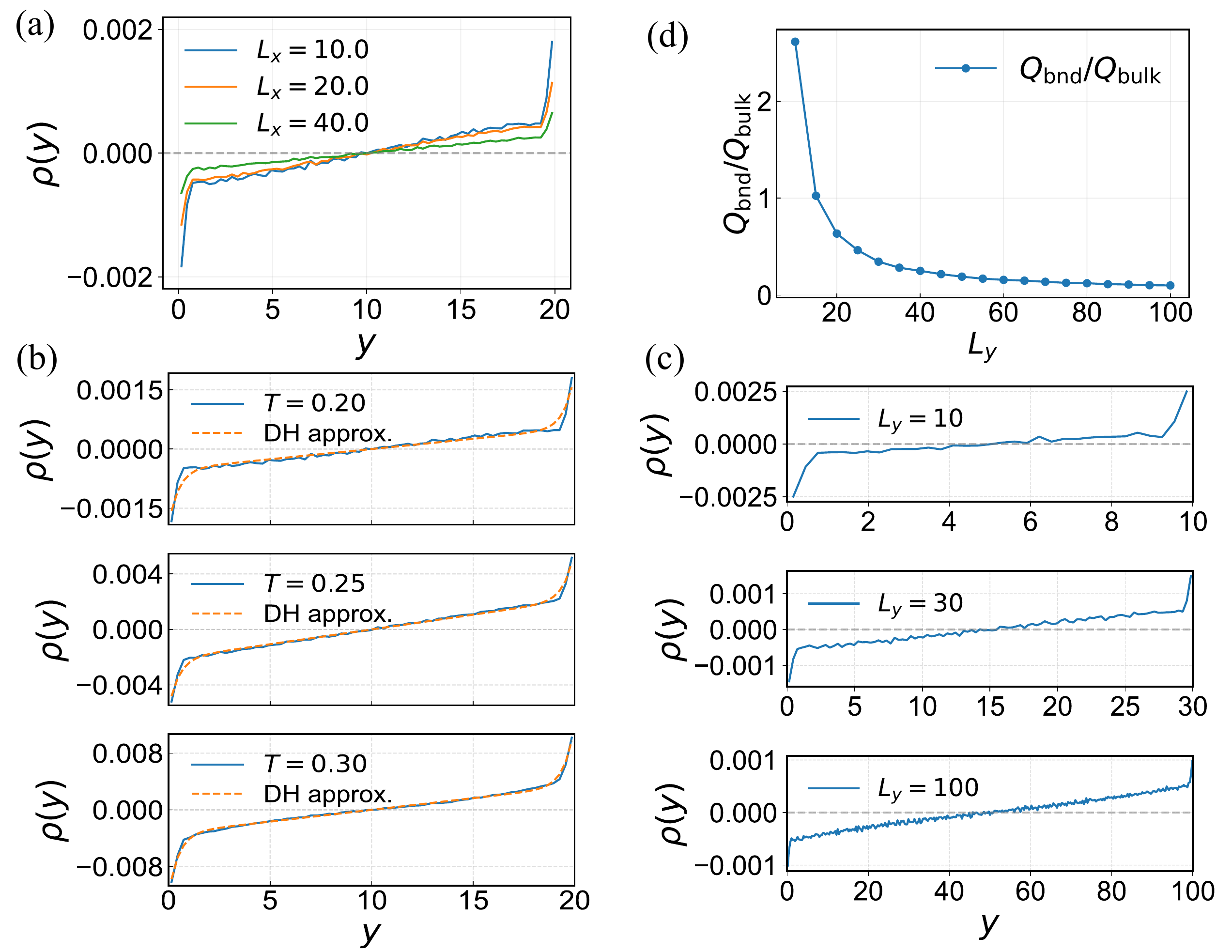}
    \caption{Induced net charge distribution.
(a)-(c) Net charge density $\rho(y)$ for different values of $L_x$, temperature $T$, and $L_y$, respectively.
Other parameters are: (a) $T=0.2$, $L_y=20$; (b) $L_x=10$, $L_y=20$; (c) $T=0.2$, $L_x=\mathrm{10}$.
The dashed lines  (b) show  $\rho(y)$ obtained
from the DH approximation, Eq.~\ref{Eq_charge_dis}.
(d) Ratio $Q_{bnd}/Q_{bulk}$ as a function of $L_y$, with $L_x=10$ and $T=0.2$.}
    \label{fig:fig2}
\end{figure}

\emph{ Monte Carlo simulation.---} 
The neutral Coulomb gas model in Eq.~\eqref{gas} is a minimal model for BKT physics and has been studied, primarily using Monte Carlo simulations \cite{kosterlitz1973, Halperin1979, Lee1992, WallinMats1997}.
The key new ingredient in our problem is the presence of opposite boundary charges at the bottom  and top edges ($y=0, L_y$), which effectively generate external electric fields in the bulk.
To screen these boundary charges and reduce the resulting electric fields, charge dipoles are induced in the bulk (see Fig.~\ref{fig:fig1}(c)).
As a consequence of this screening, positive (negative) charges are attracted toward the edge carrying negative (positive) boundary charge. We therefore perform Monte Carlo simulations to study the bulk net charge distribution $\rho(y)$ induced by the boundary charges, where $\rho(y)=\rho_{+}(y)-\rho_{-}(y)$ denotes the difference between positive and negative charge densities.
The charge distribution is uniform along the $x$ direction.
Details of the Monte Carlo method and model parameters are provided in the SM \cite{Supp}.
The BKT transition can be clearly identified in SM Fig. 1(a), where the total number of vortices increases rapidly above the BKT transition temperature.

The results are summarized in Fig.~\ref{fig:fig2}.
Without loss of generality, we set $q_{bnd}=-0.5$ at $y=0$ and $q_{bnd}=0.5$ at $y=L_y$ in the simulations.
Figure~\ref{fig:fig2}(a) shows the $L_x$ dependence of the charge distribution $\rho(y)$ at $T=0.2$ and $L_y=20$.
Note that the length unit in our simulations is roughly the scale (or a few times) of the superconducting coherence length $\xi$.
We find that $\rho(y)$ decays nearly exponentially away from the boundaries due to the presence of boundary charges.
As $L_x$ increases, the boundary charge density is reduced [see Eq.~\eqref{Eq_bc}], leading to a suppression of $\rho(y)$.
Figure~\ref{fig:fig2}(b) shows the temperature dependence of $\rho(y)$.
As expected, $\rho(y)$ increases rapidly at higher temperatures due to thermally excited positive and negative charge pairs.
Clearly, for all parameters considered, the charge distribution satisfies $\rho(y)=-\rho(-y)$, consistent with overall charge neutrality.


The charge density profile $\rho(y)$ can be obtained analytically in the regime where the charge density is sufficiently high to be treated as a continuum with Debye--H\"uckel (DH) approximation \cite{McQuarrie1976}.
In this limit, by solving the Poisson equation with boundary charge densities $\pm\sigma$ imposed at $y=0,L_y$  [for details, see SM Sec.~II], we obtain
\begin{equation}
\rho(y) \approx  -a\,\frac{\sinh[\kappa(y - L_y/2)]}{\cosh(\kappa L_y/2)} + b y,  \label{Eq_charge_dis}
\end{equation}
where $a\propto\sigma$, $b=\frac{\kappa^2 E_0}{2\pi}$, and $\kappa$ is the inverse Debye screening length.
The first term originates from the exponentially decaying electric fields near the boundaries, while the second term, linear in $y$, arises from a uniform bulk electric field $E_0$.
The analytical profile in Eq.~\eqref{Eq_charge_dis} agrees well with the numerical results [see Fig.~\ref{fig:fig2}(b)].

From the above analysis, we identify the bulk region as the region where $\rho(y)$ varies linearly with $y$, while the boundary region is defined as the region within a distance $\kappa^{-1}\sim 2$ from the edges.
As $L_y$ increases, the bulk region grows monotonically [see Fig.~\ref{fig:fig2}(c)].
To compare the charge contributions from the bulk and boundary regions, we define the boundary-induced charge
$Q_{bnd}=\int_{0}^{\kappa^{-1}}dy\,\rho(y)$
and the bulk-induced charge
$Q_{bulk}=\int_{\kappa^{-1}}^{L_y/2}dy\,\rho(y)$.
The ratio $Q_{bnd}/Q_{bulk}$ as a function of $L_y$ is shown in Fig.~\ref{fig:fig2}(d).
We find that the bulk and boundary charges become comparable at $L_y\sim 20$ for our simulation parameters.


\begin{figure}
    \centering
    \includegraphics[width=1\linewidth]{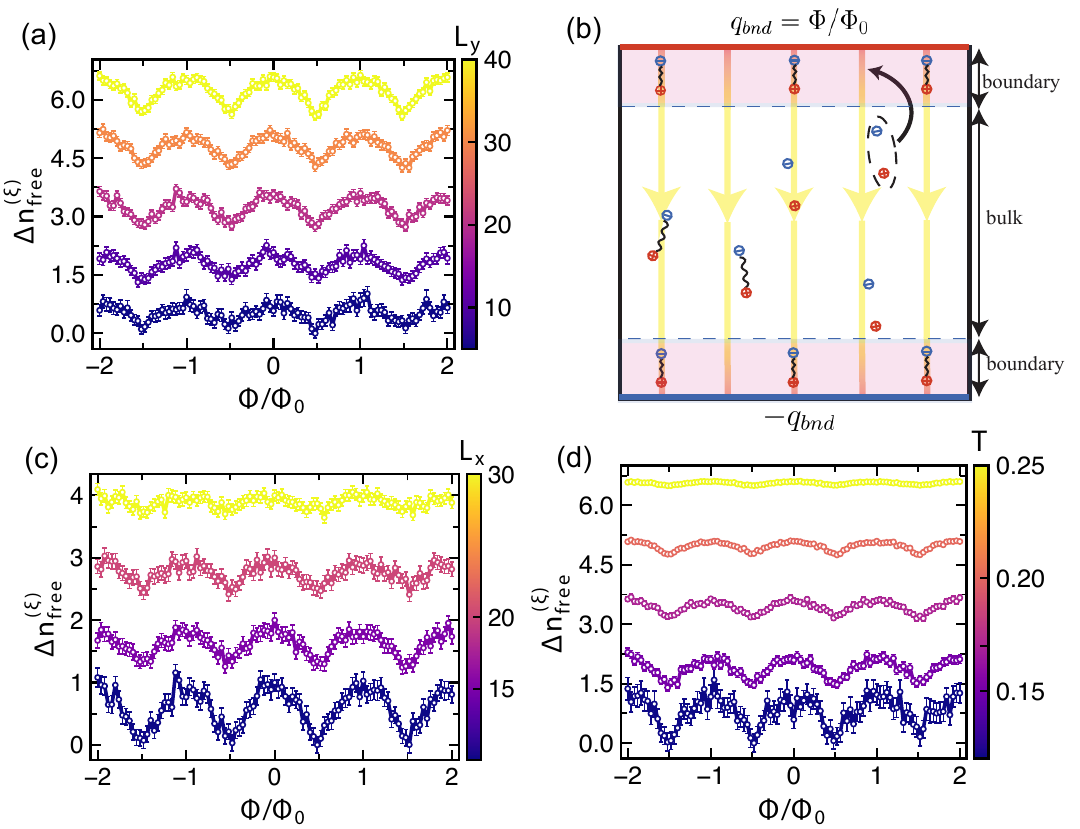}
    \caption{
(a), (c), and (d) Oscillatory free-vortex density
$\Delta n_{\mathrm{free}}^{(\xi)}$ as a function of the reduced flux
$\Phi/\Phi_{0}$. Here, $n_{\mathrm{free}}^{(\xi)}$ is inferred from the
vortex correlation length $\xi$. For clarity, the curves are vertically
offset, and the oscillatory amplitudes are normalized by the maximum
absolute value among all curves in each panel.
(a) Dependence on $L_y$ for fixed $L_x=10$ and $T=0.17$.
(c) Dependence on $L_x$ for fixed $L_y=20$ and $T=0.17$.
(d) Temperature dependence for fixed $L_x=10$ and $L_y=30$.
Error bars indicate the statistical uncertainties.
(b) Schematic illustration of vortex dynamics in the presence of
boundary charges $q_{\mathrm{bnd}}=\pm\Phi/\Phi_{0}$. The associated
electric field is stronger near the boundaries (red) and approximately
uniform and weaker in the bulk (yellow). The boundary charges attract
and pin vortex--antivortex pairs near the two edges, thereby modifying
the screening and the flux-dependent free-vortex density.
}
    \label{fig:fig3}
\end{figure}

\emph{ Half-quantum flux shift in LP oscillations.---}  As pointed out above, the boundary charge $q_{bnd}$ varies periodically with period $\Phi_0$.
Consequently, the bulk charge distribution is also expected to exhibit periodic dependence on the magnetic flux.
We therefore investigate how the free charge density $n_{free}$ varies with the magnetic flux $\Phi$ in a 2D superconducting ring.
We focus on the free charge density because, near the BKT transition, the resistance is dominated by free (i.e., unbound) vortices.
As a result, the dependence of $n_{free}$ on $\Phi$ directly corresponds to the LP oscillations of 2D superconductors near the BKT transition.   Since the resistance is proportional to $n_{\mathrm{free}}$, peaks in the curves at integer (half-integer) flux quanta correspond to the unconventional $\pi$-ring (conventional $0$-ring) behavior in the LP effect.

To obtain information on free vortices, we follow the Halperin-Nelson theory~\cite{Halperin1979} and estimate their density as
$n^{(\xi)}_{\mathrm{free}} \propto 1/\xi_v^2$,
where $\xi_v$ is the vortex correlation length. In order to make our simulation more realistic, we account for the finite-size effects associated with the cylindrical geometry by including the image-charge potential and periodic contributions in the calculation of $n^{(\xi)}_{\mathrm{free}}$. Further details are provided in the  SM Sec.~IV.

Figure~\ref{fig:fig3}(a) shows $\Delta n^{(\xi)}_{\mathrm{free}}$ as a function of $\Phi/\Phi_0$ for different values of $L_y$, exhibiting the expected oscillations with period $\Phi_0$. Here, $\Delta n^{(\xi)}_{\mathrm{free}}$ is obtained by subtracting the flux-independent background from $n^{(\xi)}_{\mathrm{free}}$ to highlight its oscillatory component. Remarkably, the system exhibits robust $\pi$-ring behavior.  As illustrated in Fig.~\ref{fig:fig3}(b), the boundary charge attracts vortex–antivortex pairs from the bulk and pins them to the boundary, thereby screening the associated electric field. As the boundary charge increases, this screening mechanism becomes stronger and further suppresses $n^{(\xi)}_{\mathrm{free}}$, giving rise to peaks at integer values of $\Phi/\Phi_0$.
 
 Finally, we examine the dependence of the LP oscillations on the ring perimeter $L_x$ and temperature $T$ [see Fig.~\ref{fig:fig3}(c) and (d)].
We find that the $\pi$-ring behavior driven by the BKT transition is highly robust and persists over a wide range of parameters.
In particular, the $\pi$-ring survives even for large aspect ratios $L_x/L_y\gg1$, although increasing $L_x$ is expected to reduce the absolute oscillation amplitude. In the Supplemental Material Fig.S2, we employ an alternative method to count the free vortices by applying the vortex-separation criterion
$|\bm{r}_i-\bm{r}_j|>r_{\mathrm{free}}$.
Although this method is less rigorous because it depends on the choice of $r_{\mathrm{free}}$, the $\pi$-ring behavior can also be clearly observed in the wide sample.




\emph{Discussion.---} In summary, we have established a theoretical framework for the LP effect induced by vortices near the BKT transition in 2D superconductors. In particular, we have shown that half-quantum flux shifts can emerge in this regime. Our proposal can be tested in clean two-dimensional superconductors with well-characterized BKT transitions. Promising candidates include atomically thin, single-crystalline transition-metal dichalcogenides.

There is currently considerable experimental interest in chiral superconductivity in two-dimensional systems, such as rhombohedral multilayer graphene \cite{Julong2025} and transition-metal dichalcogenides \cite{Wan2024, Almoalem2024}. Our theory suggests that caution is warranted when interpreting a half-quantum flux shift in LP oscillations alone as evidence of chiral superconductivity, particularly in two-dimensional superconductors. To the best of our knowledge, half-quantum-flux shifts have predominantly been observed in thin films or thin flakes of layered materials. Our results thus motivate further experimental investigation of the interplay between the observed half-quantum flux shift and BKT physics in these thin-film superconductors. If a system exhibiting half-quantum shifts is near the BKT transition, other signatures, such as enhanced low-frequency noise and a jump in the superfluid stiffness, should also emerge. Local noise magnetometry might also be useful for probing the BKT transition \cite{Demler2024}.

 Real samples often contain pinning centers, which can shift or even suppress the BKT transition \cite{Reichhardt_2017}. In particular, disorder can locally modify the vortex fugacity and superfluid stiffness, leading to a broadened or shifted BKT crossover. In sufficiently disordered samples, a sharp BKT transition may even be suppressed. Since our half-quantum flux shifted LP oscillations are closely related to vortex dynamics near the BKT regime, pinning effects can change the temperature range and magnitude of the effect. It would be interesting if the disorder might also favor the appearance of a half-quantum flux shift in some cases. We leave the study of these effects for future work.  Another interesting extension of our theory is to consider two-dimensional superconductors with periodic arrays of holes \cite{Harada_science, Berezinskii2025}, where coupling between neighboring holes may modify the LP response found for a single hole. Our framework provides a natural starting point for studying such collective interference effects and their interplay with BKT vortex dynamics.

\emph{Acknowledgment.---}	
We are very grateful for the insightful discussions with Bertrand Halperin, Leonid Glazman, and Yasuhiro Tokura.
Y.M.X. acknowledges financial support from the RIKEN Special Postdoctoral Researcher (SPDR) Program. N.N. was supported by JSPS KAKENHI Grant No.24H00197, 24H02231, and 24K00583. N.N. was also supported by the RIKEN TRIP initiative.

\bibliographystyle{apsrev4-2}
\bibliography{Reference}

\begin{thebibliography}{36}%
\makeatletter
\providecommand \@ifxundefined [1]{%
 \@ifx{#1\undefined}
}%
\providecommand \@ifnum [1]{%
 \ifnum #1\expandafter \@firstoftwo
 \else \expandafter \@secondoftwo
 \fi
}%
\providecommand \@ifx [1]{%
 \ifx #1\expandafter \@firstoftwo
 \else \expandafter \@secondoftwo
 \fi
}%
\providecommand \natexlab [1]{#1}%
\providecommand \enquote  [1]{``#1''}%
\providecommand \bibnamefont  [1]{#1}%
\providecommand \bibfnamefont [1]{#1}%
\providecommand \citenamefont [1]{#1}%
\providecommand \href@noop [0]{\@secondoftwo}%
\providecommand \href [0]{\begingroup \@sanitize@url \@href}%
\providecommand \@href[1]{\@@startlink{#1}\@@href}%
\providecommand \@@href[1]{\endgroup#1\@@endlink}%
\providecommand \@sanitize@url [0]{\catcode `\\12\catcode `\$12\catcode `\&12\catcode `\#12\catcode `\^12\catcode `\_12\catcode `\%12\relax}%
\providecommand \@@startlink[1]{}%
\providecommand \@@endlink[0]{}%
\providecommand \url  [0]{\begingroup\@sanitize@url \@url }%
\providecommand \@url [1]{\endgroup\@href {#1}{\urlprefix }}%
\providecommand \urlprefix  [0]{URL }%
\providecommand \Eprint [0]{\href }%
\providecommand \doibase [0]{https://doi.org/}%
\providecommand \selectlanguage [0]{\@gobble}%
\providecommand \bibinfo  [0]{\@secondoftwo}%
\providecommand \bibfield  [0]{\@secondoftwo}%
\providecommand \translation [1]{[#1]}%
\providecommand \BibitemOpen [0]{}%
\providecommand \bibitemStop [0]{}%
\providecommand \bibitemNoStop [0]{.\EOS\space}%
\providecommand \EOS [0]{\spacefactor3000\relax}%
\providecommand \BibitemShut  [1]{\csname bibitem#1\endcsname}%
\let\auto@bib@innerbib\@empty
\bibitem [{\citenamefont {Deaver}\ and\ \citenamefont {Fairbank}(1961)}]{Deaver1961}%
  \BibitemOpen
  \bibfield  {author} {\bibinfo {author} {\bibfnamefont {B.~S.}\ \bibnamefont {Deaver}}\ and\ \bibinfo {author} {\bibfnamefont {W.~M.}\ \bibnamefont {Fairbank}},\ }\href {https://doi.org/10.1103/PhysRevLett.7.43} {\bibfield  {journal} {\bibinfo  {journal} {Phys. Rev. Lett.}\ }\textbf {\bibinfo {volume} {7}},\ \bibinfo {pages} {43} (\bibinfo {year} {1961})}\BibitemShut {NoStop}%
\bibitem [{\citenamefont {Doll}\ and\ \citenamefont {N\"abauer}(1961)}]{Doll1961}%
  \BibitemOpen
  \bibfield  {author} {\bibinfo {author} {\bibfnamefont {R.}~\bibnamefont {Doll}}\ and\ \bibinfo {author} {\bibfnamefont {M.}~\bibnamefont {N\"abauer}},\ }\href {https://doi.org/10.1103/PhysRevLett.7.51} {\bibfield  {journal} {\bibinfo  {journal} {Phys. Rev. Lett.}\ }\textbf {\bibinfo {volume} {7}},\ \bibinfo {pages} {51} (\bibinfo {year} {1961})}\BibitemShut {NoStop}%
\bibitem [{\citenamefont {Little}\ and\ \citenamefont {Parks}(1962)}]{Little1962}%
  \BibitemOpen
  \bibfield  {author} {\bibinfo {author} {\bibfnamefont {W.~A.}\ \bibnamefont {Little}}\ and\ \bibinfo {author} {\bibfnamefont {R.~D.}\ \bibnamefont {Parks}},\ }\href {https://doi.org/10.1103/PhysRevLett.9.9} {\bibfield  {journal} {\bibinfo  {journal} {Phys. Rev. Lett.}\ }\textbf {\bibinfo {volume} {9}},\ \bibinfo {pages} {9} (\bibinfo {year} {1962})}\BibitemShut {NoStop}%
\bibitem [{\citenamefont {Parks}\ and\ \citenamefont {Little}(1964)}]{Park1964}%
  \BibitemOpen
  \bibfield  {author} {\bibinfo {author} {\bibfnamefont {R.~D.}\ \bibnamefont {Parks}}\ and\ \bibinfo {author} {\bibfnamefont {W.~A.}\ \bibnamefont {Little}},\ }\href {https://doi.org/10.1103/PhysRev.133.A97} {\bibfield  {journal} {\bibinfo  {journal} {Phys. Rev.}\ }\textbf {\bibinfo {volume} {133}},\ \bibinfo {pages} {A97} (\bibinfo {year} {1964})}\BibitemShut {NoStop}%
\bibitem [{\citenamefont {Tinkham}(2004)}]{tinkham2004}%
  \BibitemOpen
  \bibfield  {author} {\bibinfo {author} {\bibfnamefont {M.}~\bibnamefont {Tinkham}},\ }\href@noop {} {\emph {\bibinfo {title} {Introduction to superconductivity}}}\ (\bibinfo  {publisher} {Courier Corporation},\ \bibinfo {year} {2004})\BibitemShut {NoStop}%
\bibitem [{\citenamefont {Grosso}\ and\ \citenamefont {Parravicini}(2013)}]{grosso2013solid}%
  \BibitemOpen
  \bibfield  {author} {\bibinfo {author} {\bibfnamefont {G.}~\bibnamefont {Grosso}}\ and\ \bibinfo {author} {\bibfnamefont {G.~P.}\ \bibnamefont {Parravicini}},\ }\href@noop {} {\emph {\bibinfo {title} {Solid state physics}}}\ (\bibinfo  {publisher} {Academic press},\ \bibinfo {year} {2013})\BibitemShut {NoStop}%
\bibitem [{\citenamefont {Tsuei}\ and\ \citenamefont {Kirtley}(2000)}]{Tsuei2000}%
  \BibitemOpen
  \bibfield  {author} {\bibinfo {author} {\bibfnamefont {C.~C.}\ \bibnamefont {Tsuei}}\ and\ \bibinfo {author} {\bibfnamefont {J.~R.}\ \bibnamefont {Kirtley}},\ }\href {https://doi.org/10.1103/RevModPhys.72.969} {\bibfield  {journal} {\bibinfo  {journal} {Rev. Mod. Phys.}\ }\textbf {\bibinfo {volume} {72}},\ \bibinfo {pages} {969} (\bibinfo {year} {2000})}\BibitemShut {NoStop}%
\bibitem [{\citenamefont {Sigrist}(2005)}]{Manfred2005}%
  \BibitemOpen
  \bibfield  {author} {\bibinfo {author} {\bibfnamefont {M.}~\bibnamefont {Sigrist}},\ }\href {https://doi.org/10.1063/1.2080350} {\bibfield  {journal} {\bibinfo  {journal} {AIP Conference Proceedings}\ }\textbf {\bibinfo {volume} {789}},\ \bibinfo {pages} {165} (\bibinfo {year} {2005})}\BibitemShut {NoStop}%
\bibitem [{\citenamefont {Geshkenbein}\ \emph {et~al.}(1987)\citenamefont {Geshkenbein}, \citenamefont {Larkin},\ and\ \citenamefont {Barone}}]{GLB1987}%
  \BibitemOpen
  \bibfield  {author} {\bibinfo {author} {\bibfnamefont {V.~B.}\ \bibnamefont {Geshkenbein}}, \bibinfo {author} {\bibfnamefont {A.~I.}\ \bibnamefont {Larkin}},\ and\ \bibinfo {author} {\bibfnamefont {A.}~\bibnamefont {Barone}},\ }\href {https://doi.org/10.1103/PhysRevB.36.235} {\bibfield  {journal} {\bibinfo  {journal} {Phys. Rev. B}\ }\textbf {\bibinfo {volume} {36}},\ \bibinfo {pages} {235} (\bibinfo {year} {1987})}\BibitemShut {NoStop}%
\bibitem [{\citenamefont {Chung}\ \emph {et~al.}(2007)\citenamefont {Chung}, \citenamefont {Bluhm},\ and\ \citenamefont {Kim}}]{Kim2007}%
  \BibitemOpen
  \bibfield  {author} {\bibinfo {author} {\bibfnamefont {S.~B.}\ \bibnamefont {Chung}}, \bibinfo {author} {\bibfnamefont {H.}~\bibnamefont {Bluhm}},\ and\ \bibinfo {author} {\bibfnamefont {E.-A.}\ \bibnamefont {Kim}},\ }\href {https://doi.org/10.1103/PhysRevLett.99.197002} {\bibfield  {journal} {\bibinfo  {journal} {Phys. Rev. Lett.}\ }\textbf {\bibinfo {volume} {99}},\ \bibinfo {pages} {197002} (\bibinfo {year} {2007})}\BibitemShut {NoStop}%
\bibitem [{\citenamefont {Vakaryuk}\ and\ \citenamefont {Leggett}(2009)}]{Victor2009}%
  \BibitemOpen
  \bibfield  {author} {\bibinfo {author} {\bibfnamefont {V.}~\bibnamefont {Vakaryuk}}\ and\ \bibinfo {author} {\bibfnamefont {A.~J.}\ \bibnamefont {Leggett}},\ }\href {https://doi.org/10.1103/PhysRevLett.103.057003} {\bibfield  {journal} {\bibinfo  {journal} {Phys. Rev. Lett.}\ }\textbf {\bibinfo {volume} {103}},\ \bibinfo {pages} {057003} (\bibinfo {year} {2009})}\BibitemShut {NoStop}%
\bibitem [{\citenamefont {Salomaa}\ and\ \citenamefont {Volovik}(1985)}]{Volovik1985}%
  \BibitemOpen
  \bibfield  {author} {\bibinfo {author} {\bibfnamefont {M.~M.}\ \bibnamefont {Salomaa}}\ and\ \bibinfo {author} {\bibfnamefont {G.~E.}\ \bibnamefont {Volovik}},\ }\href {https://doi.org/10.1103/PhysRevLett.55.1184} {\bibfield  {journal} {\bibinfo  {journal} {Phys. Rev. Lett.}\ }\textbf {\bibinfo {volume} {55}},\ \bibinfo {pages} {1184} (\bibinfo {year} {1985})}\BibitemShut {NoStop}%
\bibitem [{\citenamefont {Li}\ \emph {et~al.}(2019)\citenamefont {Li}, \citenamefont {Xu}, \citenamefont {Lee}, \citenamefont {Chu},\ and\ \citenamefont {Chien}}]{Yufan2019}%
  \BibitemOpen
  \bibfield  {author} {\bibinfo {author} {\bibfnamefont {Y.}~\bibnamefont {Li}}, \bibinfo {author} {\bibfnamefont {X.}~\bibnamefont {Xu}}, \bibinfo {author} {\bibfnamefont {M.-H.}\ \bibnamefont {Lee}}, \bibinfo {author} {\bibfnamefont {M.-W.}\ \bibnamefont {Chu}},\ and\ \bibinfo {author} {\bibfnamefont {C.~L.}\ \bibnamefont {Chien}},\ }\href {https://doi.org/10.1126/science.aau6539} {\bibfield  {journal} {\bibinfo  {journal} {Science}\ }\textbf {\bibinfo {volume} {366}},\ \bibinfo {pages} {238} (\bibinfo {year} {2019})}\BibitemShut {NoStop}%
\bibitem [{\citenamefont {Xu}\ \emph {et~al.}(2020)\citenamefont {Xu}, \citenamefont {Li},\ and\ \citenamefont {Chien}}]{Xiaoying2020}%
  \BibitemOpen
  \bibfield  {author} {\bibinfo {author} {\bibfnamefont {X.}~\bibnamefont {Xu}}, \bibinfo {author} {\bibfnamefont {Y.}~\bibnamefont {Li}},\ and\ \bibinfo {author} {\bibfnamefont {C.~L.}\ \bibnamefont {Chien}},\ }\href {https://doi.org/10.1103/PhysRevLett.124.167001} {\bibfield  {journal} {\bibinfo  {journal} {Phys. Rev. Lett.}\ }\textbf {\bibinfo {volume} {124}},\ \bibinfo {pages} {167001} (\bibinfo {year} {2020})}\BibitemShut {NoStop}%
\bibitem [{\citenamefont {Wan}\ \emph {et~al.}(2024)\citenamefont {Wan}, \citenamefont {Qiu}, \citenamefont {Ren}, \citenamefont {Qian}, \citenamefont {Li}, \citenamefont {Xu}, \citenamefont {Zhou}, \citenamefont {Zhou}, \citenamefont {Zhou}, \citenamefont {Wang}, \citenamefont {Yang}, \citenamefont {Sofer}, \citenamefont {Huang}, \citenamefont {Wang},\ and\ \citenamefont {Duan}}]{Wan2024}%
  \BibitemOpen
  \bibfield  {author} {\bibinfo {author} {\bibfnamefont {Z.}~\bibnamefont {Wan}}, \bibinfo {author} {\bibfnamefont {G.}~\bibnamefont {Qiu}}, \bibinfo {author} {\bibfnamefont {H.}~\bibnamefont {Ren}}, \bibinfo {author} {\bibfnamefont {Q.}~\bibnamefont {Qian}}, \bibinfo {author} {\bibfnamefont {Y.}~\bibnamefont {Li}}, \bibinfo {author} {\bibfnamefont {D.}~\bibnamefont {Xu}}, \bibinfo {author} {\bibfnamefont {J.}~\bibnamefont {Zhou}}, \bibinfo {author} {\bibfnamefont {J.}~\bibnamefont {Zhou}}, \bibinfo {author} {\bibfnamefont {B.}~\bibnamefont {Zhou}}, \bibinfo {author} {\bibfnamefont {L.}~\bibnamefont {Wang}}, \bibinfo {author} {\bibfnamefont {T.-H.}\ \bibnamefont {Yang}}, \bibinfo {author} {\bibfnamefont {Z.}~\bibnamefont {Sofer}}, \bibinfo {author} {\bibfnamefont {Y.}~\bibnamefont {Huang}}, \bibinfo {author} {\bibfnamefont {K.~L.}\ \bibnamefont {Wang}},\ and\ \bibinfo {author} {\bibfnamefont {X.}~\bibnamefont {Duan}},\ }\href {https://doi.org/10.1038/s41586-024-07625-4} {\bibfield  {journal} {\bibinfo
  {journal} {Nature}\ }\textbf {\bibinfo {volume} {632}},\ \bibinfo {pages} {69} (\bibinfo {year} {2024})}\BibitemShut {NoStop}%
\bibitem [{\citenamefont {Almoalem}\ \emph {et~al.}(2024)\citenamefont {Almoalem}, \citenamefont {Feldman}, \citenamefont {Mangel}, \citenamefont {Shlafman}, \citenamefont {Yaish}, \citenamefont {Fischer}, \citenamefont {Moshe}, \citenamefont {Ruhman},\ and\ \citenamefont {Kanigel}}]{Almoalem2024}%
  \BibitemOpen
  \bibfield  {author} {\bibinfo {author} {\bibfnamefont {A.}~\bibnamefont {Almoalem}}, \bibinfo {author} {\bibfnamefont {I.}~\bibnamefont {Feldman}}, \bibinfo {author} {\bibfnamefont {I.}~\bibnamefont {Mangel}}, \bibinfo {author} {\bibfnamefont {M.}~\bibnamefont {Shlafman}}, \bibinfo {author} {\bibfnamefont {Y.~E.}\ \bibnamefont {Yaish}}, \bibinfo {author} {\bibfnamefont {M.~H.}\ \bibnamefont {Fischer}}, \bibinfo {author} {\bibfnamefont {M.}~\bibnamefont {Moshe}}, \bibinfo {author} {\bibfnamefont {J.}~\bibnamefont {Ruhman}},\ and\ \bibinfo {author} {\bibfnamefont {A.}~\bibnamefont {Kanigel}},\ }\href {https://doi.org/10.1038/s41467-024-48260-x} {\bibfield  {journal} {\bibinfo  {journal} {Nature Communications}\ }\textbf {\bibinfo {volume} {15}},\ \bibinfo {pages} {4623} (\bibinfo {year} {2024})}\BibitemShut {NoStop}%
\bibitem [{\citenamefont {Tokuda}\ \emph {et~al.}(2025)\citenamefont {Tokuda}, \citenamefont {Matsumoto}, \citenamefont {Maeda}, \citenamefont {Higashihara}, \citenamefont {Nakao}, \citenamefont {Watanabe}, \citenamefont {Lee}, \citenamefont {Nakamura}, \citenamefont {Maeda}, \citenamefont {Jiang}, \citenamefont {Yue}, \citenamefont {Narita}, \citenamefont {Aoyama}, \citenamefont {Mizushima}, \citenamefont {ichiro Ohe}, \citenamefont {Ono}, \citenamefont {Jin}, \citenamefont {Kobayashi},\ and\ \citenamefont {Niimi}}]{Tokuda2025}%
  \BibitemOpen
  \bibfield  {author} {\bibinfo {author} {\bibfnamefont {M.}~\bibnamefont {Tokuda}}, \bibinfo {author} {\bibfnamefont {F.}~\bibnamefont {Matsumoto}}, \bibinfo {author} {\bibfnamefont {N.}~\bibnamefont {Maeda}}, \bibinfo {author} {\bibfnamefont {T.}~\bibnamefont {Higashihara}}, \bibinfo {author} {\bibfnamefont {M.}~\bibnamefont {Nakao}}, \bibinfo {author} {\bibfnamefont {M.}~\bibnamefont {Watanabe}}, \bibinfo {author} {\bibfnamefont {S.}~\bibnamefont {Lee}}, \bibinfo {author} {\bibfnamefont {R.}~\bibnamefont {Nakamura}}, \bibinfo {author} {\bibfnamefont {M.}~\bibnamefont {Maeda}}, \bibinfo {author} {\bibfnamefont {N.}~\bibnamefont {Jiang}}, \bibinfo {author} {\bibfnamefont {D.}~\bibnamefont {Yue}}, \bibinfo {author} {\bibfnamefont {H.}~\bibnamefont {Narita}}, \bibinfo {author} {\bibfnamefont {K.}~\bibnamefont {Aoyama}}, \bibinfo {author} {\bibfnamefont {T.}~\bibnamefont {Mizushima}}, \bibinfo {author} {\bibfnamefont {J.}~\bibnamefont {ichiro Ohe}}, \bibinfo {author} {\bibfnamefont {T.}~\bibnamefont {Ono}},
  \bibinfo {author} {\bibfnamefont {X.}~\bibnamefont {Jin}}, \bibinfo {author} {\bibfnamefont {K.}~\bibnamefont {Kobayashi}},\ and\ \bibinfo {author} {\bibfnamefont {Y.}~\bibnamefont {Niimi}},\ }\href {https://doi.org/10.1126/sciadv.adw6625} {\bibfield  {journal} {\bibinfo  {journal} {Science Advances}\ }\textbf {\bibinfo {volume} {11}},\ \bibinfo {pages} {eadw6625} (\bibinfo {year} {2025})}\BibitemShut {NoStop}%
\bibitem [{\citenamefont {{Wang}}\ \emph {et~al.}(2025)\citenamefont {{Wang}}, \citenamefont {{Maccari}}, \citenamefont {{Feng}}, \citenamefont {{Wu}}, \citenamefont {{Peng}}, \citenamefont {{Tuen Law}}, \citenamefont {{Zhao}}, \citenamefont {{Szabo}}, \citenamefont {{Schnyder}}, \citenamefont {{Kang}}, \citenamefont {{Wu}}, \citenamefont {{Liu}}, \citenamefont {{Fu}}, \citenamefont {{Fischer}}, \citenamefont {{Sigrist}}, \citenamefont {{Yu}},\ and\ \citenamefont {{Lin}}}]{benchuan2025}%
  \BibitemOpen
  \bibfield  {author} {\bibinfo {author} {\bibfnamefont {S.}~\bibnamefont {{Wang}}}, \bibinfo {author} {\bibfnamefont {I.}~\bibnamefont {{Maccari}}}, \bibinfo {author} {\bibfnamefont {X.}~\bibnamefont {{Feng}}}, \bibinfo {author} {\bibfnamefont {Z.-N.}\ \bibnamefont {{Wu}}}, \bibinfo {author} {\bibfnamefont {J.-P.}\ \bibnamefont {{Peng}}}, \bibinfo {author} {\bibfnamefont {K.}~\bibnamefont {{Tuen Law}}}, \bibinfo {author} {\bibfnamefont {Y.~X.}\ \bibnamefont {{Zhao}}}, \bibinfo {author} {\bibfnamefont {A.}~\bibnamefont {{Szabo}}}, \bibinfo {author} {\bibfnamefont {A.}~\bibnamefont {{Schnyder}}}, \bibinfo {author} {\bibfnamefont {N.}~\bibnamefont {{Kang}}}, \bibinfo {author} {\bibfnamefont {X.-S.}\ \bibnamefont {{Wu}}}, \bibinfo {author} {\bibfnamefont {J.}~\bibnamefont {{Liu}}}, \bibinfo {author} {\bibfnamefont {X.}~\bibnamefont {{Fu}}}, \bibinfo {author} {\bibfnamefont {M.~H.}\ \bibnamefont {{Fischer}}}, \bibinfo {author} {\bibfnamefont {M.}~\bibnamefont {{Sigrist}}}, \bibinfo {author} {\bibfnamefont
  {D.}~\bibnamefont {{Yu}}},\ and\ \bibinfo {author} {\bibfnamefont {B.-C.}\ \bibnamefont {{Lin}}},\ }\href {https://doi.org/10.48550/arXiv.2512.10010} {\bibfield  {journal} {\bibinfo  {journal} {arXiv e-prints}\ ,\ \bibinfo {eid} {arXiv:2512.10010}} (\bibinfo {year} {2025})}\BibitemShut {NoStop}%
\bibitem [{\citenamefont {Zhang}\ \emph {et~al.}()\citenamefont {Zhang}, \citenamefont {Xie}, \citenamefont {Liu}, \citenamefont {Fang}, \citenamefont {Zhang}, \citenamefont {Zou}, \citenamefont {Zhang}, \citenamefont {Jia}, \citenamefont {Chen}, \citenamefont {Ma}, \citenamefont {Zhao}, \citenamefont {Loh}, \citenamefont {Huang},\ and\ \citenamefont {Xiu}}]{ZhangTRSB}%
  \BibitemOpen
  \bibfield  {author} {\bibinfo {author} {\bibfnamefont {E.}~\bibnamefont {Zhang}}, \bibinfo {author} {\bibfnamefont {Y.-M.}\ \bibnamefont {Xie}}, \bibinfo {author} {\bibfnamefont {S.}~\bibnamefont {Liu}}, \bibinfo {author} {\bibfnamefont {Y.}~\bibnamefont {Fang}}, \bibinfo {author} {\bibfnamefont {J.}~\bibnamefont {Zhang}}, \bibinfo {author} {\bibfnamefont {Y.-C.}\ \bibnamefont {Zou}}, \bibinfo {author} {\bibfnamefont {Y.}~\bibnamefont {Zhang}}, \bibinfo {author} {\bibfnamefont {Z.}~\bibnamefont {Jia}}, \bibinfo {author} {\bibfnamefont {J.}~\bibnamefont {Chen}}, \bibinfo {author} {\bibfnamefont {Q.}~\bibnamefont {Ma}}, \bibinfo {author} {\bibfnamefont {W.}~\bibnamefont {Zhao}}, \bibinfo {author} {\bibfnamefont {K.~P.}\ \bibnamefont {Loh}}, \bibinfo {author} {\bibfnamefont {F.}~\bibnamefont {Huang}},\ and\ \bibinfo {author} {\bibfnamefont {F.}~\bibnamefont {Xiu}},\ }\bibinfo {note} {manuscript submitted}\BibitemShut {NoStop}%
\bibitem [{\citenamefont {Aoyama}(2022)}]{Aoyama2022}%
  \BibitemOpen
  \bibfield  {author} {\bibinfo {author} {\bibfnamefont {K.}~\bibnamefont {Aoyama}},\ }\href {https://doi.org/10.1103/PhysRevB.106.L060502} {\bibfield  {journal} {\bibinfo  {journal} {Phys. Rev. B}\ }\textbf {\bibinfo {volume} {106}},\ \bibinfo {pages} {L060502} (\bibinfo {year} {2022})}\BibitemShut {NoStop}%
\bibitem [{\citenamefont {Fischer}\ \emph {et~al.}(2023)\citenamefont {Fischer}, \citenamefont {Lee},\ and\ \citenamefont {Ruhman}}]{Lee2023}%
  \BibitemOpen
  \bibfield  {author} {\bibinfo {author} {\bibfnamefont {M.~H.}\ \bibnamefont {Fischer}}, \bibinfo {author} {\bibfnamefont {P.~A.}\ \bibnamefont {Lee}},\ and\ \bibinfo {author} {\bibfnamefont {J.}~\bibnamefont {Ruhman}},\ }\href {https://doi.org/10.1103/PhysRevB.108.L180505} {\bibfield  {journal} {\bibinfo  {journal} {Phys. Rev. B}\ }\textbf {\bibinfo {volume} {108}},\ \bibinfo {pages} {L180505} (\bibinfo {year} {2023})}\BibitemShut {NoStop}%
\bibitem [{\citenamefont {Hua}\ \emph {et~al.}(2023)\citenamefont {Hua}, \citenamefont {Dumitrescu},\ and\ \citenamefont {Hal\'asz}}]{Hua2023}%
  \BibitemOpen
  \bibfield  {author} {\bibinfo {author} {\bibfnamefont {C.}~\bibnamefont {Hua}}, \bibinfo {author} {\bibfnamefont {E.}~\bibnamefont {Dumitrescu}},\ and\ \bibinfo {author} {\bibfnamefont {G.~B.}\ \bibnamefont {Hal\'asz}},\ }\href {https://doi.org/10.1103/PhysRevB.107.214503} {\bibfield  {journal} {\bibinfo  {journal} {Phys. Rev. B}\ }\textbf {\bibinfo {volume} {107}},\ \bibinfo {pages} {214503} (\bibinfo {year} {2023})}\BibitemShut {NoStop}%
\bibitem [{\citenamefont {{Zhang}}\ and\ \citenamefont {{Li}}(2025)}]{LiYi2025}%
  \BibitemOpen
  \bibfield  {author} {\bibinfo {author} {\bibfnamefont {J.}~\bibnamefont {{Zhang}}}\ and\ \bibinfo {author} {\bibfnamefont {Y.}~\bibnamefont {{Li}}},\ }\href {https://doi.org/10.48550/arXiv.2509.24581} {\bibfield  {journal} {\bibinfo  {journal} {arXiv e-prints}\ ,\ \bibinfo {eid} {arXiv:2509.24581}} (\bibinfo {year} {2025})}\BibitemShut {NoStop}%
\bibitem [{\citenamefont {Nagaosa}(2013)}]{nagaosa2013quantum}%
  \BibitemOpen
  \bibfield  {author} {\bibinfo {author} {\bibfnamefont {N.}~\bibnamefont {Nagaosa}},\ }\href@noop {} {\emph {\bibinfo {title} {Quantum field theory in condensed matter physics}}}\ (\bibinfo  {publisher} {Springer Science \& Business Media},\ \bibinfo {year} {2013})\BibitemShut {NoStop}%
\bibitem [{Sup()}]{Supp}%
  \BibitemOpen
  \href@noop {} {}\bibinfo {note} {See the Supplementary Material for (i) field theory details for Little-Parks oscillations near BKT transitions, (ii)Monte Carlo Simulation Method, (iii) induced charge density distribution under the DH approximation.}\BibitemShut {Stop}%
\bibitem [{\citenamefont {Fisher}\ and\ \citenamefont {Lee}(1989)}]{Fisher}%
  \BibitemOpen
  \bibfield  {author} {\bibinfo {author} {\bibfnamefont {M.~P.~A.}\ \bibnamefont {Fisher}}\ and\ \bibinfo {author} {\bibfnamefont {D.~H.}\ \bibnamefont {Lee}},\ }\href {https://doi.org/10.1103/PhysRevB.39.2756} {\bibfield  {journal} {\bibinfo  {journal} {Phys. Rev. B}\ }\textbf {\bibinfo {volume} {39}},\ \bibinfo {pages} {2756} (\bibinfo {year} {1989})}\BibitemShut {NoStop}%
\bibitem [{\citenamefont {Kosterlitz}\ and\ \citenamefont {Thouless}(1973)}]{kosterlitz1973}%
  \BibitemOpen
  \bibfield  {author} {\bibinfo {author} {\bibfnamefont {J.~M.}\ \bibnamefont {Kosterlitz}}\ and\ \bibinfo {author} {\bibfnamefont {D.~J.}\ \bibnamefont {Thouless}},\ }\href@noop {} {\bibfield  {journal} {\bibinfo  {journal} {Journal of Physics C: Solid State Physics}\ }\textbf {\bibinfo {volume} {6}},\ \bibinfo {pages} {1181} (\bibinfo {year} {1973})}\BibitemShut {NoStop}%
\bibitem [{\citenamefont {Halperin}\ and\ \citenamefont {Nelson}(1979)}]{Halperin1979}%
  \BibitemOpen
  \bibfield  {author} {\bibinfo {author} {\bibfnamefont {B.~I.}\ \bibnamefont {Halperin}}\ and\ \bibinfo {author} {\bibfnamefont {D.~R.}\ \bibnamefont {Nelson}},\ }\href {https://doi.org/10.1007/BF00116988} {\bibfield  {journal} {\bibinfo  {journal} {Journal of Low Temperature Physics}\ }\textbf {\bibinfo {volume} {36}},\ \bibinfo {pages} {599} (\bibinfo {year} {1979})}\BibitemShut {NoStop}%
\bibitem [{\citenamefont {Lee}\ and\ \citenamefont {Teitel}(1992)}]{Lee1992}%
  \BibitemOpen
  \bibfield  {author} {\bibinfo {author} {\bibfnamefont {J.-R.}\ \bibnamefont {Lee}}\ and\ \bibinfo {author} {\bibfnamefont {S.}~\bibnamefont {Teitel}},\ }\href {https://doi.org/10.1103/PhysRevB.46.3247} {\bibfield  {journal} {\bibinfo  {journal} {Phys. Rev. B}\ }\textbf {\bibinfo {volume} {46}},\ \bibinfo {pages} {3247} (\bibinfo {year} {1992})}\BibitemShut {NoStop}%
\bibitem [{\citenamefont {Lidmar}\ and\ \citenamefont {Wallin}(1997)}]{WallinMats1997}%
  \BibitemOpen
  \bibfield  {author} {\bibinfo {author} {\bibfnamefont {J.}~\bibnamefont {Lidmar}}\ and\ \bibinfo {author} {\bibfnamefont {M.}~\bibnamefont {Wallin}},\ }\href {https://doi.org/10.1103/PhysRevB.55.522} {\bibfield  {journal} {\bibinfo  {journal} {Phys. Rev. B}\ }\textbf {\bibinfo {volume} {55}},\ \bibinfo {pages} {522} (\bibinfo {year} {1997})}\BibitemShut {NoStop}%
\bibitem [{\citenamefont {McQuarrie}(1976)}]{McQuarrie1976}%
  \BibitemOpen
  \bibfield  {author} {\bibinfo {author} {\bibfnamefont {D.~A.}\ \bibnamefont {McQuarrie}},\ }\href@noop {} {\emph {\bibinfo {title} {Statistical Mechanics}}}\ (\bibinfo  {publisher} {Harper \& Row},\ \bibinfo {address} {New York},\ \bibinfo {year} {1976})\BibitemShut {NoStop}%
\bibitem [{\citenamefont {Han}\ \emph {et~al.}(2025)\citenamefont {Han}, \citenamefont {Lu}, \citenamefont {Hadjri}, \citenamefont {Shi}, \citenamefont {Wu}, \citenamefont {Xu}, \citenamefont {Yao}, \citenamefont {Cotten}, \citenamefont {Sharifi~Sedeh}, \citenamefont {Weldeyesus}, \citenamefont {Yang}, \citenamefont {Seo}, \citenamefont {Ye}, \citenamefont {Zhou}, \citenamefont {Liu}, \citenamefont {Shi}, \citenamefont {Hua}, \citenamefont {Watanabe}, \citenamefont {Taniguchi}, \citenamefont {Xiong}, \citenamefont {Zumb{\"u}hl}, \citenamefont {Fu},\ and\ \citenamefont {Ju}}]{Julong2025}%
  \BibitemOpen
  \bibfield  {author} {\bibinfo {author} {\bibfnamefont {T.}~\bibnamefont {Han}}, \bibinfo {author} {\bibfnamefont {Z.}~\bibnamefont {Lu}}, \bibinfo {author} {\bibfnamefont {Z.}~\bibnamefont {Hadjri}}, \bibinfo {author} {\bibfnamefont {L.}~\bibnamefont {Shi}}, \bibinfo {author} {\bibfnamefont {Z.}~\bibnamefont {Wu}}, \bibinfo {author} {\bibfnamefont {W.}~\bibnamefont {Xu}}, \bibinfo {author} {\bibfnamefont {Y.}~\bibnamefont {Yao}}, \bibinfo {author} {\bibfnamefont {A.~A.}\ \bibnamefont {Cotten}}, \bibinfo {author} {\bibfnamefont {O.}~\bibnamefont {Sharifi~Sedeh}}, \bibinfo {author} {\bibfnamefont {H.}~\bibnamefont {Weldeyesus}}, \bibinfo {author} {\bibfnamefont {J.}~\bibnamefont {Yang}}, \bibinfo {author} {\bibfnamefont {J.}~\bibnamefont {Seo}}, \bibinfo {author} {\bibfnamefont {S.}~\bibnamefont {Ye}}, \bibinfo {author} {\bibfnamefont {M.}~\bibnamefont {Zhou}}, \bibinfo {author} {\bibfnamefont {H.}~\bibnamefont {Liu}}, \bibinfo {author} {\bibfnamefont {G.}~\bibnamefont {Shi}}, \bibinfo {author} {\bibfnamefont
  {Z.}~\bibnamefont {Hua}}, \bibinfo {author} {\bibfnamefont {K.}~\bibnamefont {Watanabe}}, \bibinfo {author} {\bibfnamefont {T.}~\bibnamefont {Taniguchi}}, \bibinfo {author} {\bibfnamefont {P.}~\bibnamefont {Xiong}}, \bibinfo {author} {\bibfnamefont {D.~M.}\ \bibnamefont {Zumb{\"u}hl}}, \bibinfo {author} {\bibfnamefont {L.}~\bibnamefont {Fu}},\ and\ \bibinfo {author} {\bibfnamefont {L.}~\bibnamefont {Ju}},\ }\href {https://doi.org/10.1038/s41586-025-09169-7} {\bibfield  {journal} {\bibinfo  {journal} {Nature}\ }\textbf {\bibinfo {volume} {643}},\ \bibinfo {pages} {654} (\bibinfo {year} {2025})}\BibitemShut {NoStop}%
\bibitem [{\citenamefont {Curtis}\ \emph {et~al.}(2024)\citenamefont {Curtis}, \citenamefont {Maksimovic}, \citenamefont {Poniatowski}, \citenamefont {Yacoby}, \citenamefont {Halperin}, \citenamefont {Narang},\ and\ \citenamefont {Demler}}]{Demler2024}%
  \BibitemOpen
  \bibfield  {author} {\bibinfo {author} {\bibfnamefont {J.~B.}\ \bibnamefont {Curtis}}, \bibinfo {author} {\bibfnamefont {N.}~\bibnamefont {Maksimovic}}, \bibinfo {author} {\bibfnamefont {N.~R.}\ \bibnamefont {Poniatowski}}, \bibinfo {author} {\bibfnamefont {A.}~\bibnamefont {Yacoby}}, \bibinfo {author} {\bibfnamefont {B.}~\bibnamefont {Halperin}}, \bibinfo {author} {\bibfnamefont {P.}~\bibnamefont {Narang}},\ and\ \bibinfo {author} {\bibfnamefont {E.}~\bibnamefont {Demler}},\ }\href {https://doi.org/10.1103/PhysRevB.110.144518} {\bibfield  {journal} {\bibinfo  {journal} {Phys. Rev. B}\ }\textbf {\bibinfo {volume} {110}},\ \bibinfo {pages} {144518} (\bibinfo {year} {2024})}\BibitemShut {NoStop}%
\bibitem [{\citenamefont {Reichhardt}\ and\ \citenamefont {Olson~Reichhardt}(2016)}]{Reichhardt_2017}%
  \BibitemOpen
  \bibfield  {author} {\bibinfo {author} {\bibfnamefont {C.}~\bibnamefont {Reichhardt}}\ and\ \bibinfo {author} {\bibfnamefont {C.~J.}\ \bibnamefont {Olson~Reichhardt}},\ }\href {https://doi.org/10.1088/1361-6633/80/2/026501} {\bibfield  {journal} {\bibinfo  {journal} {Reports on Progress in Physics}\ }\textbf {\bibinfo {volume} {80}},\ \bibinfo {pages} {026501} (\bibinfo {year} {2016})}\BibitemShut {NoStop}%
\bibitem [{\citenamefont {Harada}\ \emph {et~al.}(1996)\citenamefont {Harada}, \citenamefont {Kamimura}, \citenamefont {Kasai}, \citenamefont {Matsuda}, \citenamefont {Tonomura},\ and\ \citenamefont {Moshchalkov}}]{Harada_science}%
  \BibitemOpen
  \bibfield  {author} {\bibinfo {author} {\bibfnamefont {K.}~\bibnamefont {Harada}}, \bibinfo {author} {\bibfnamefont {O.}~\bibnamefont {Kamimura}}, \bibinfo {author} {\bibfnamefont {H.}~\bibnamefont {Kasai}}, \bibinfo {author} {\bibfnamefont {T.}~\bibnamefont {Matsuda}}, \bibinfo {author} {\bibfnamefont {A.}~\bibnamefont {Tonomura}},\ and\ \bibinfo {author} {\bibfnamefont {V.~V.}\ \bibnamefont {Moshchalkov}},\ }\href {https://doi.org/10.1126/science.274.5290.1167} {\bibfield  {journal} {\bibinfo  {journal} {Science}\ }\textbf {\bibinfo {volume} {274}},\ \bibinfo {pages} {1167} (\bibinfo {year} {1996})}\BibitemShut {NoStop}%
\bibitem [{\citenamefont {Verma}\ \emph {et~al.}(2025)\citenamefont {Verma}, \citenamefont {Vedin}, \citenamefont {Jesudasan}, \citenamefont {Lidmar}, \citenamefont {Maccari},\ and\ \citenamefont {Bose}}]{Berezinskii2025}%
  \BibitemOpen
  \bibfield  {author} {\bibinfo {author} {\bibfnamefont {A.}~\bibnamefont {Verma}}, \bibinfo {author} {\bibfnamefont {R.}~\bibnamefont {Vedin}}, \bibinfo {author} {\bibfnamefont {J.}~\bibnamefont {Jesudasan}}, \bibinfo {author} {\bibfnamefont {J.}~\bibnamefont {Lidmar}}, \bibinfo {author} {\bibfnamefont {I.}~\bibnamefont {Maccari}},\ and\ \bibinfo {author} {\bibfnamefont {S.}~\bibnamefont {Bose}},\ }\href {https://doi.org/10.1103/p1t4-hst3} {\bibfield  {journal} {\bibinfo  {journal} {Phys. Rev. B}\ }\textbf {\bibinfo {volume} {112}},\ \bibinfo {pages} {L220501} (\bibinfo {year} {2025})}\BibitemShut {NoStop}%
\end{thebibliography}%

 \clearpage
	
	\onecolumngrid
	\begin{center}
		\textbf{\large Supplementary Material for  \lq\lq{} Theory of Little--Parks oscillations  by vortices in two-dimensional superconductors \rq\rq{}}\\[.2cm]
		Ying-Ming Xie$^{1,2}$ and Nato Nagaosa$^{2,3}$ \\[.1cm]
		{\itshape ${}^1$   Institute of Condensed Matter Physics, School of Physics and Astronomy, \\Shanghai Jiao Tong University, Shanghai, China}\\{\itshape ${}^2$    RIKEN Center for Emergent Matter Science (CEMS), Wako, Saitama 351-0198, Japan}\\
        {\itshape ${}^3$  Fundamental Quantum Science Program (FQSP), TRIP Headquarters, RIKEN, Wako 351-0198, Japan}
	\end{center}
	\setcounter{equation}{0}
	\setcounter{section}{0}
	\setcounter{figure}{0}
	\setcounter{table}{0}
	\setcounter{page}{1}
	\renewcommand{\theequation}{S\arabic{equation}}
	\renewcommand{\thetable}{S\arabic{table}}
	\renewcommand{\thesection}{\Roman{section}}
	\renewcommand{\thefigure}{S\arabic{figure}}
	\renewcommand{\bibnumfmt}[1]{[S#1]}
	\renewcommand{\citenumfont}[1]{#1}
	\makeatletter

\onecolumngrid

\maketitle

\setcounter{tocdepth}{1} 
\tableofcontents

\section{ Field theory details for Little-Parks oscillations near BKT transitions }

\subsection{Field theory near the BKT transition for 2D superconductors}

The action is given by
\begin{equation}
S=\int d^3\tilde{x}\,\frac{J}{2}\left(\partial_{\mu}\theta+2eA_{\mu}\right)^2 .
\end{equation}
All notation used in the Supplementary Material follows the definitions in the main text.
To proceed, we decompose the phase field into a smooth part and a vortex-core contribution,
\[
\theta=\theta_0+\theta_v.
\]
Applying a Hubbard--Stratonovich transformation to decouple the quadratic term yields
\begin{equation}
\exp\!\left[-\int d^3\tilde{x}\,\frac{J}{2}(\partial_{\mu}\theta+2eA_{\mu})^2\right]
\propto
\int D J_{\mu}\,
\exp\!\left[-\int d^3\tilde{x}
\left(\frac{J_{\mu}^2}{2J}
+iJ_{\mu}(\partial_{\mu}\theta_0+\partial_{\mu}\theta_v+2eA_{\mu})\right)\right].
\end{equation}
The partition function can therefore be written as
\begin{equation}
Z=\int D\theta_0\,D\theta_v\,D J_{\mu}\,
\exp\!\left[-\int d^3\tilde{x}
\left(\frac{J_{\mu}^2}{2J}
+iJ_{\mu}(\partial_{\mu}\theta_0+\partial_{\mu}\theta_v+2eA_{\mu})\right)\right].
\end{equation}

We now isolate the contribution involving the smooth phase field \(\theta_0\),
\begin{equation}
S[\theta_0]
=\int d^3\tilde{x}\, iJ_{\mu}\partial_{\mu}\theta_0
=-\int d^3\tilde{x}\, i(\partial_{\mu}J_{\mu})\,\theta_0,
\end{equation}
where the second equality follows from integration by parts, assuming vanishing boundary terms.
Performing the functional integral over \(\theta_0\) enforces the conservation of the auxiliary current,
\begin{equation}
\int D\theta_0\, e^{-S[\theta_0]}
=\int D\theta_0\,
\exp\!\left(\int d^3\tilde{x}\, i(\partial_{\mu}J_{\mu})\,\theta_0\right)
\propto \delta[\partial_{\mu}J_{\mu}] .
\end{equation}
After integrating out \(\theta_0\), the partition function reduces to
\begin{equation}
Z=\int D\theta_v\,D J_{\mu}\,
\delta[\partial_{\mu}J_{\mu}]\,
\exp\!\left[-\int d^3\tilde{x}
\left(\frac{J_{\mu}^2}{2J}
+iJ_{\mu}(\partial_{\mu}\theta_v+2eA_{\mu})\right)\right].
\end{equation}

The consdervation of supercurrent $\partial_{\mu} J_{\mu}=0$, resulting in
\begin{equation}
J_{\mu}= \frac{1}{2\pi}\epsilon_{\mu\nu\lambda}\partial_{\nu}a_{\lambda}=\frac{1}{2\pi} \nabla_{3D}\times a_{3D}.
\end{equation}

\begin{equation}
    \int d^3\tilde{x} i J_{\mu}(x) \partial_{\mu}\theta_{v}=\int d^3\tilde{x} i\frac{1}{2\pi}\epsilon_{\mu\nu\lambda}\partial_{\nu} a_{\lambda} \partial_{\mu}\theta_{v}=\int d^3\tilde{x} i\frac{1}{2\pi}\epsilon_{\nu\mu\lambda} a_{\lambda} \partial_{\nu}\partial_{\mu}\theta_{v}=\int d^3\tilde{x} ia_{\lambda}j_{\lambda}
\end{equation}

\begin{equation}
   2e \int d^3\tilde{x} \frac{i}{2\pi}\epsilon_{\mu\nu\lambda}\partial_{\nu} a_{\lambda} A_{\mu}=\frac{e}{\pi}\int d^3\tilde{x} i a_{\lambda} B_{\lambda}
\end{equation}

The resulting effective action is
\begin{equation}
\tilde{S}'
= A \sum_{i} \int ds_i
+ \int d^3x\,\frac{(\nabla_{3D}\times a_{3D})^2}{8\pi^2J}
+ i\int d^3x\, a_{3D}\cdot\left(j_{3D}+\frac{e}{\pi}B_{3D}\right).
\label{total_ac}
\end{equation}
The first term on the right-hand side represents the energy cost associated with vortex cores, with an action $A$ per unit length multiplied by the length of the world line $\int ds_i$ of the $i$th vortex.

In the static two-dimensional limit, the imaginary-time integral reduces to
$\int_0^{\beta} d\tau = \beta.$
Moreover,
\begin{equation}
(\nabla_{3D}\times a_{3D})^2 = F_{\mu}F_{\mu},
\end{equation}
where
$F_{\mu}=\epsilon_{\mu\nu\lambda}\partial_{\nu}a_{\lambda}.$
Explicitly, the components are
\begin{eqnarray}
F_0 &=& \epsilon_{0ij}\partial_i a_j
= \partial_x a_y - \partial_y a_x, \\
F_i &=& \epsilon_{i0j}\partial_0 a_j + \epsilon_{ij0}\partial_j a_0
= -\epsilon_{ij}\partial_0 a_j + \epsilon_{ij}\partial_j a_0 .
\end{eqnarray}
The external magnetic field is given by
$B_{3D}=B_0
=\epsilon_{0\nu\mu}\partial_{\nu}A_{\mu}
=\nabla_{2D}\times\bm{A}
= B_z\,\hat{\bm{z}} .$

In the static limit, we may set $a_i=0$ since the spatial components do not couple to the vortex density, and $\partial_0 a_i=0$. Consequently, the theory reduces to the mapping
\begin{equation}
a_{3D} \mapsto a_0(\bm{r}), \qquad
j_{\mu} \mapsto j_0=\rho_v(\bm{r}), \qquad
B_{3D} \mapsto B_0 = B_z .
\end{equation}
The effective action then becomes
\begin{equation}
\tilde{S}'
= \beta \int d^2\bm{r}
\left\{
A|\rho_v(\bm{r})|
+ i a_0(\bm{r})\!\left(\rho_v(\bm{r})+\frac{e}{\pi}B_z\right)
+ \frac{|\nabla a_0(\bm{r})|^2}{8\pi^2J}
\right\}.
\end{equation}
This action describes a two-dimensional Coulomb gas.

\subsection{Map to a Coulomb gas problem}

Let us take the vortex charge density to be
\begin{equation}
\rho_{v}=\sum_{i} q_i\delta(\bm{r}-\bm{X}_i).
\end{equation}
Here $q_i=\pm 1$ corresponds to a vortex or an antivortex, and $\bm{X}_i$ denotes the position of the $i$-th vortex core. We now integrate out the scalar field $a_0(\bm{r})$ in the zero-field limit ($B_z=0$).

The action is
\begin{eqnarray}
\tilde{S}'[a_0,\rho_v]&&=\int d^2r \left[\frac{1}{8\pi^2 J}(\nabla a_0)^2+ia_0(\bm{r})\rho_v(\bm{r})\right]\\
&&=\int d^2r \left[-\frac{1}{8\pi^2 J}a_0\nabla^2 a_0+ia_0(\bm{r})\rho_v(\bm{r})\right],
\end{eqnarray}
where the second line follows from integration by parts.

Using the standard Gaussian integral formula
$$
\int D\phi\, e^{-\frac{1}{2}(\phi, K\phi)+(J,\phi)}
\propto \exp\!\left[\frac{1}{2}(J, K^{-1}J)\right],
$$
we obtain
\begin{equation}
Z\propto \exp\!\left[-\frac{1}{2}(\rho_v, K^{-1}\rho_v)\right]
=\exp\!\left[-\frac{1}{2}\int d^2r\, d^2r'\,
\rho_v(\bm{r})\, V(\bm{r}-\bm{r}')\, \rho_v(\bm{r}')\right].
\end{equation}

Here the interaction kernel $V(\bm{r})$ is defined through the relation
$K V(\bm{r})=\delta(\bm{r})$, which implies
\begin{equation}
\nabla^2 V(\bm{r})= - 4\pi^2 J\,\delta(\bm{r}). \label{possion}
\end{equation}

In Fourier space, this yields $V(\bm{k})=4\pi^2 J/k^2$. We also make use of the identity
\begin{equation}
 \int \frac{d^2k}{(2\pi)^2} \frac{e^{i \bm{k}\cdot \bm{r}}}{k^2}
 =-\frac{1}{2\pi}\ln \frac{r}{r_c},
\end{equation}
where $r_c$ is a short-distance cutoff.

The effective action therefore becomes
\begin{eqnarray}
\tilde{S}_{\mathrm{eff}}&&=\frac{1}{2}\int d^2r\, d^2r'\,
\rho_v(\bm{r})\, V(\bm{r}-\bm{r}')\, \rho_v(\bm{r}')\nonumber\\
&&=\frac{1}{2}\sum_{i\neq j} q_i q_j\, V(\bm{X}_{ij})
+ E_c\sum_{i} q_i^2,
\end{eqnarray}
where $\bm{X}_{ij}=\bm{X}_i-\bm{X}_j$, and the vortex core energy is given by
$E_c=\frac{1}{2}V(0)$. The interaction potential between vortices is
\begin{equation}
V(\bm{X}_{ij})=-2\pi J \ln \frac{|\bm{X}_{ij}|}{r_c}.
\end{equation}

\subsection{Dual-field theory description of the Little-Parks effect near BKT transition}
We consider a superconducting cylinder of radius $R$, threaded by a magnetic flux
$\Phi$ through its center. Inside the sample, the magnetic field satisfies
$B_z=0$, while the vector potential $\bm{A}$ remains nonzero, corresponding to
an Aharonov--Bohm configuration. The central question is how the vector
potential $\bm{A}$ affects the Kosterlitz--Thouless  transition.

From the dual action, the coupling to the external field is given by
\begin{equation}
    \int d^3x \frac{ie}{\pi} \partial_{\nu} a_{\lambda} A_{\mu}
    =\frac{ i e}{\pi}\int d^3x (\nabla_{3D}\times a_{3D})\cdot \bm{A}_{3D}.
\end{equation}

Using the vector identity
\begin{equation}
\nabla\cdot (\bm{c}\times \bm{d})
=\bm{d}\cdot (\nabla\times \bm{c})
-\bm{c}\cdot (\nabla\times \bm{d}),
\end{equation}
and replacing $\bm{c}=\bm{a}_{3D}$, $\bm{d}=\bm{A}_{3D}$, together with the
condition $\nabla_{3D}\times \bm{A}_{3D}=0$ inside the sample region, we obtain
in the static limit
\begin{eqnarray}
    \int d^3x \frac{ie}{\pi} \partial_{\nu} a_{\lambda} A_{\mu}
    &&= \frac{ ie}{\pi}\int d^3 x\,\nabla \cdot
    ( \bm{a}_{3D} \times \bm{A}_{3D})\nonumber\\
    &&= \frac{ie \beta}{\pi} \int _{\partial V}
    d^2\bm{r}\cdot ( \bm{a}_{3D} \times \bm{A}_{3D}).
\end{eqnarray}

The boundaries of the cylinder are located at $y=0$ and $y=L_y$, and
\begin{equation}
\bm{a}_{3D}\times \bm{A}_{3D}
= (a_x A_y- a_y A_x,\; a_y A_0- a_0 A_y,\; a_0 A_x- a_x A_0).
\end{equation}

In the static limit $A_0=0$, this reduces to
\begin{equation}
  \int d^3x \frac{ie}{\pi} \partial_{\nu} a_{\lambda} A_{\mu}
  = \frac{ie \beta}{\pi}
  \int d^2 r\, a_0 A_x \bigl[\delta (y-L_y)-\delta (y)\bigr].
\end{equation}

Upon unwrapping the cylindrical geometry into Cartesian coordinates $(x,y)$,
the vector potential becomes $A_x=\Phi/(2\pi R)$. The effective field-theoretic
description of the Little--Parks experiment near the BKT transition is then
given by
\begin{equation}
S_{BKT}= \beta \int d^2r
\left[
i a_0(\bm{r})\bigl(\rho_v(\bm{r})+ \rho_{v}^{(A)}(\bm{r})\bigr)
+ \frac{|\nabla a_0(\bm{r})|^2}{8\pi^2J}
\right].
\end{equation}

Here the boundary-induced vortex charge density is
\begin{equation}
    \rho_{v}^{(A)}(\bm{r})
    =\frac{e\Phi}{2\pi^2R}\bigl[\delta(y-L_y)-\delta(y)\bigr].
\end{equation}
We have neglected the vortex core energy contribution (the first term in
Eq.~\eqref{total_ac}), since it merely shifts the overall normalization of the
partition function.

Because $\hbar=1$, the superconducting flux quantum is $\Phi_0=\pi/e$.
For flux values $\Phi=n\Phi_0$, the vector potential can be absorbed as a large vortex trapped inside the cylinder. We therefore redefine the
boundary vortex charge density as
\begin{equation}
    \rho_{v}^{(A)}
    =\frac{q_{bnd}}{L_x}\bigl[\delta(y-L_y)-\delta(y)\bigr],
\end{equation}
where
\begin{equation}
    q_{bnd}=\left(\frac{\Phi}{\Phi_0}-n\right),
\end{equation}
and $n$ is an integer chosen such that
$|\Phi/\Phi_0-n|<\frac{1}{2}$.

\section{Induced charge density distribution under the DH approximation}

The charge density in the bulk can be expressed as
\begin{eqnarray}
\rho(\mathbf{r})
&=& n_{+} e^{-\beta q_{+}\phi(\mathbf{r})}
    - n_{-} e^{+\beta q_{-}\phi(\mathbf{r})} \nonumber\\
&=& -2n_{0}\sinh[\beta \phi(\mathbf{r})],
\end{eqnarray}
where $\phi(\mathbf{r})$ denotes the electrostatic potential field, and we have assumed charge symmetry with $n_{+} = n_{-} = n_{0}$ and $q_{+} = -q_{-} = 1$.

The electrostatic potential satisfies the two-dimensional Poisson equation,
\begin{equation}
\nabla^{2}\phi(\mathbf{r}) = -2\pi \rho(\mathbf{r}).
\end{equation}
Substituting the expression for $\rho(\mathbf{r})$ yields
\begin{equation}
\nabla^{2}\phi(\mathbf{r}) = 4\pi n_{0}\sinh[\beta \phi(\mathbf{r})].
\end{equation}

In the weak-potential limit ($\beta\phi \ll 1$), we linearize the hyperbolic sine function, $\sinh(\beta\phi) \simeq \beta\phi$.
This leads to the linearized Debye--H\"uckel (DH) approximation,
\begin{equation}
\nabla^{2}\phi(\mathbf{r}) = \kappa^{2}\phi(\mathbf{r}), \label{Eq_72}
\end{equation}
where the inverse Debye length is defined as $\kappa=\lambda_D^{-1}$

Before solving this differential equation explicitly, we can expect that the solution of this equation near the boundary decays exponentially, $\phi(r) \sim e^{-r / \lambda_{D}}$, with the \textit{Debye screening length}
\begin{equation}
\lambda_{D} = \frac{1}{\kappa}
= \frac{1}{\sqrt{4\pi \beta n_{0}}}.
\end{equation}
Therefore, in the linearized regime, the potential and the induced charge density both decay over the characteristic length scale $\lambda_{D}$.
Consequently, the localization length of the boundary-induced charge distribution is governed by the Debye length $\lambda_{D}$. According to Ref.~\cite{Halperin1979},
\begin{equation}
    n_0\propto \frac{1}{\xi_{+}^2}
\end{equation}
where the characteristic length scale controlling the vortex density is
$\xi_{+} = \xi_0 \exp\!\left( \frac{b}{\sqrt{t}} \right)$,
with
$t = \frac{T - T_{\mathrm{BKT}}}{T_{\mathrm{BKT}}}$.
Hence,
\begin{equation}
    \lambda_D \propto \xi_{+}= \xi_0 \exp[b/\sqrt{t}].
\end{equation}

Let us perform a preliminary estimate of the order of magnitude of the vortex density.
In practice, $b \sim 1$ \cite{kosterlitz1973}.
Using $T_{\mathrm{BKT}} = 0.18$, we estimate
$n_0(T = 0.2) \sim 10^{-4}\text{--}10^{-3},$
which is consistent with the numerical results presented in the main text (see Fig.~3).

Now we solve Eq.~\eqref{Eq_72} in a more accurate manner. 
In the cylindrical geometry, the system is periodic along the \(x\)-direction. 
The equation then simplifies to 
\begin{equation}
\frac{\partial^2 \phi}{\partial y^2} = \kappa^2 \phi(y).
\end{equation}

According to Gauss’s law, \(\bm{\hat{n}}\cdot \bm{E} = \sigma / \varepsilon\), 
where \(\bm{\hat{n}}\) is the surface normal vector, \(\sigma\) is the surface charge density, 
and \(\varepsilon\) is the dielectric permittivity. 
Since the surface normals and charge densities at \(y = 0\) and \(y = L\) are opposite with a boundary charge of  $+\sigma$ and $-\sigma$ respectively, 
the boundary conditions are
\begin{equation}
\phi'(0) = \phi'(L) = \frac{\sigma}{\varepsilon}.
\end{equation}

We take the ansatz 
\begin{equation}
\phi(y) = \phi_{\mathrm{DH}}(y) - E_0 y,
\end{equation}
where \(E_0 = \sigma_{\mathrm{net}} / \varepsilon\) characterizes the residual uniform 
electric field that remains after screening. 
By solving the differential equation with the above boundary conditions, we obtain
\begin{equation}
\phi(y) = \left(\frac{\sigma}{\varepsilon \kappa} + \frac{E_0}{\kappa}\right)
\frac{\sinh[\kappa(y - L/2)]}{\cosh(\kappa L/2)} - E_0 y,
\end{equation}

The carrier density is then given by
\begin{equation}
\rho(y) \approx -\frac{\kappa^2}{2\pi} \phi(y)
= -a\,\frac{\sinh[\kappa(y - L/2)]}{\cosh(\kappa L/2)} + b y,
\end{equation}
where \(a = \frac{\kappa}{2\pi}\left(\frac{\sigma}{\varepsilon} + E_0\right)\) 
and \(b = \frac{\kappa^2 E_0}{2\pi}\).

\section{Monte Carlo Simulation Method}

\subsection{Monte Carlo   simulation for 2D neutral Coulomb gas}

We simulate a two-dimensional Coulomb gas representing vortex--antivortex excitations of the XY model near the Kosterlitz--Thouless (KT) transition. Each configuration consists of integer charges $q_i = \pm 1$ interacting via a logarithmic potential 
\[
E = -\sum_{i<j} q_i q_j \ln\!\left(\frac{\bm{X}_{ij}}{r_c}\right),
\]
with a short-range cutoff $r_{\min}$. The Coulomb gas is charge neutral with $\sum_{i} q_i=0$. The system is defined on a square of size $L \times L$.

Monte Carlo sampling is performed using a Metropolis algorithm that combines (i) displacement moves of existing vortices, (ii) creation of vortex--antivortex pairs, and (iii) annihilation of existing pairs.  Each move is accepted with probability $\min[1, e^{-\beta \Delta E}]$, with energy differences $\Delta E$ computed efficiently using Numba-accelerated routines. The creation and destruction moves include the appropriate phase-space and chemical potential factors to satisfy detailed balance. We followed the Monte Carlo simulation method in Ref.~\cite{WallinMats1997}. But for completeness, let us explain the Monte Carlo updates in more detail below.

We simulate a two-dimensional Coulomb gas in the grand-canonical ensemble at inverse temperature $\beta$ and chemical potential $\mu$ per particle.
A configuration $C$ consists of $N$ charges $\{(q_i,\mathbf r_i)\}$ with $q_i=\pm1$ and statistical weight
\begin{equation}
\mathcal{P}(C)\propto \exp\!\left[-\beta\left(E(C)-\mu N\right)\right].
\end{equation}
Charge neutrality is enforced by allowing only the creation and destruction of neutral $(+,-)$ pairs.
For pair creation, starting from a configuration $C$ with $N=2n_{\rm old}$ particles, a trial move $C\to C'$ is proposed as follows:
a $+1$ charge is inserted uniformly over the system area $V=L_xL_y$, followed by the insertion of a $-1$ charge uniformly within a disk of radius $d$.
The corresponding proposal area is $\Omega\equiv V_b=\pi d^2$.
The move is rejected immediately if any hard-core constraint is violated (e.g., if interparticle distances are smaller than $r_{\min}$).
The trial probability density for pair creation is therefore
$t_{\rm cre}(C\to C')=\frac{1}{V\,\Omega}.$
For the reverse move $C'\to C$, the system contains
$n_{\rm new}=\frac{N+2}{2}=n_{\rm old}+1 $
neutral pairs.
One pair is chosen uniformly at random and removed, yielding the trial probability
$t_{\rm des}(C'\to C)=\frac{1}{n_{\rm new}}.$
Let $\Delta E=E(C')-E(C)$ denote the total energy change associated with inserting the proposed pair, including all interaction and boundary contributions.
Detailed balance requires
\begin{equation}
\mathcal{P}(C)\,t(C\to C')\,\tilde{a}(C\to C')
=
\mathcal{P}(C')\,t(C'\to C)\,\tilde{a}(C'\to C),
\end{equation}
where $\tilde{a}$ denotes the acceptance probability.
Using the Metropolis choice $a=\min(1,R)$, the acceptance ratio for pair creation is
$R_{\rm cre}
=
\frac{V\,\Omega}{n_{\rm old}+1}
\exp\!\left[-\beta\left(\Delta E - 2\mu\right)\right].$
The creation move is therefore accepted with probability
$a_{\rm cre}
=
\min\!\left[
1,\;
\frac{V\,\Omega}{n_{\rm old}+1}
\exp\!\left(-\beta(\Delta E-2\mu)\right)
\right]. $
The corresponding destruction move uses the inverse ratio,
$a_{\rm des}=\min\!\left(1, R_{\rm cre}^{-1}\right).$
After a thermalization period of approximately $2\times10^4$ sweeps, measurements are collected every few sweeps over about $10\times10^4$ updates. For convenience, the key Monte Carlo parameters are summarized in Table~\ref{tab:key_MC_params}. {The BKT transition, determined from the temperature dependence of the total vortex number, is shown in Fig.~\ref{Fig_S1}(a).} 

{
To estimate the free-vortex density more directly, we use the relation
$n_{\mathrm{free}}^{(\xi)} \propto 1/\xi_{v}^2$, where $\xi_{v}$ is the vortex
correlation length \cite{Halperin1979}. Near the BKT transition,
$C(\Delta r) \equiv C(r-r') \propto e^{-\Delta r/\xi_v}$, so that the
logarithm of the correlation function is linear in separation:
$\ln\!\left|C(\Delta r)\right| = \mathrm{constant} - \Delta r/\xi_v$.
The calculated correlation function is shown in Fig.~\ref{Fig_S1}(b), and the inverse
correlation length is obtained from the slope of the linear regime.
By varying the flux, we then obtain the Little--Parks oscillations of
$n_{\mathrm{free}}^{(\xi)}$ for several values of $L_y$.}

{We also display another way to evaluate $n_{free}$ in Fig.~\ref{fig:figs2}. In this mehtod, we dermine the $n_{free}$ via the counting method with $r_{free}$. Specifically, for each vortex at $\mathbf{r}_i$, we identify the nearest vortex of opposite charge at $\mathbf{r}_j$. The vortex is classified as free if $|\mathbf{r}_i-\mathbf{r}_j|>r_{\mathrm{free}}$. It is worth noting that  this method is less rigorous because it depends on the choice of $r_{\mathrm{free}}$. However, as shown in Fig.~\ref{fig:figs2}, the $\pi$-ring behavior can also be robustly observed in the wide samples.}

\begin{table}
\centering
\caption{Key parameters used in the Monte Carlo simulation of the 2D neutral Coulomb gas.}
\begin{tabular}{lcl}
\hline
\textbf{Parameter} & \textbf{Symbol / Value} & \textbf{Description} \\
\hline
Chemical potential & $\mu = -0.80$ & Controls vortex pair density \\
Core cutoff & $r_{\min} = 0.35$ & Hard-core exclusion radius \\
Interaction scale & $r_c = 0.94$ & Reference distance in $\log(r/r_c)$ interaction \\
Maximum separation& $d=1$&  the maximum separation for a local pair\\
Thermalization & $N_{\mathrm{therm}} = 2.0\times10^4$ & Number of equilibration sweeps \\
Measurement & $N_{\mathrm{meas}} = 10.0\times10^4$ & Number of measurement sweeps \\
Measurement interval & $N_{\mathrm{skip}} = 20$ & Sweeps between successive measurements \\
\hline
\end{tabular}
\label{tab:key_MC_params}
\end{table}
\subsection{Boundary potential from uniform line charges}

We consider a two--dimensional Coulomb gas confined between two open boundaries at 
$y=0$ and $y=L_y$, with periodic boundary conditions along $x$.  
Each boundary carries a uniform line charge of opposite sign, 
$-\lambda$ at $y=0$ and $+\lambda$ at $y=L_y$,  
producing a one--body potential acting on mobile charges in the interior.

The electrostatic potential at position $(x,y)$ due to a uniform line charge 
with density $\lambda$ along $x' \in [0,L_x]$ at height $y_0$ is
\begin{equation}
V_{\rm line}(x,y)
= -\lambda \int_0^{L_x} \ln\!\sqrt{(x-x')^2+(y-y_0)^2}\,dx'.
\end{equation}
Since the line is translationally invariant in $x$, the result depends only on 
the vertical separation $a = |y-y_0|$.  
Evaluating the integral gives
\begin{equation}
\Gamma(a,L_x)
\equiv \int_0^{L_x} \ln\!\sqrt{u^2+a^2}\,du
= \tfrac{1}{2} L_x \ln(L_x^2+a^2) + a \arctan\!\frac{L_x}{a} - L_x. \label{eq:Phi_exact}
\end{equation}
Up to an additive constant (which cancels in $\Delta E$), 
the potential of a single uniform line is therefore $V_{\rm line}(y) = -\lambda\,\Gamma(|y-y_0|,L_x)$.

\begin{figure}
    \centering
    \includegraphics[width=0.8\linewidth]{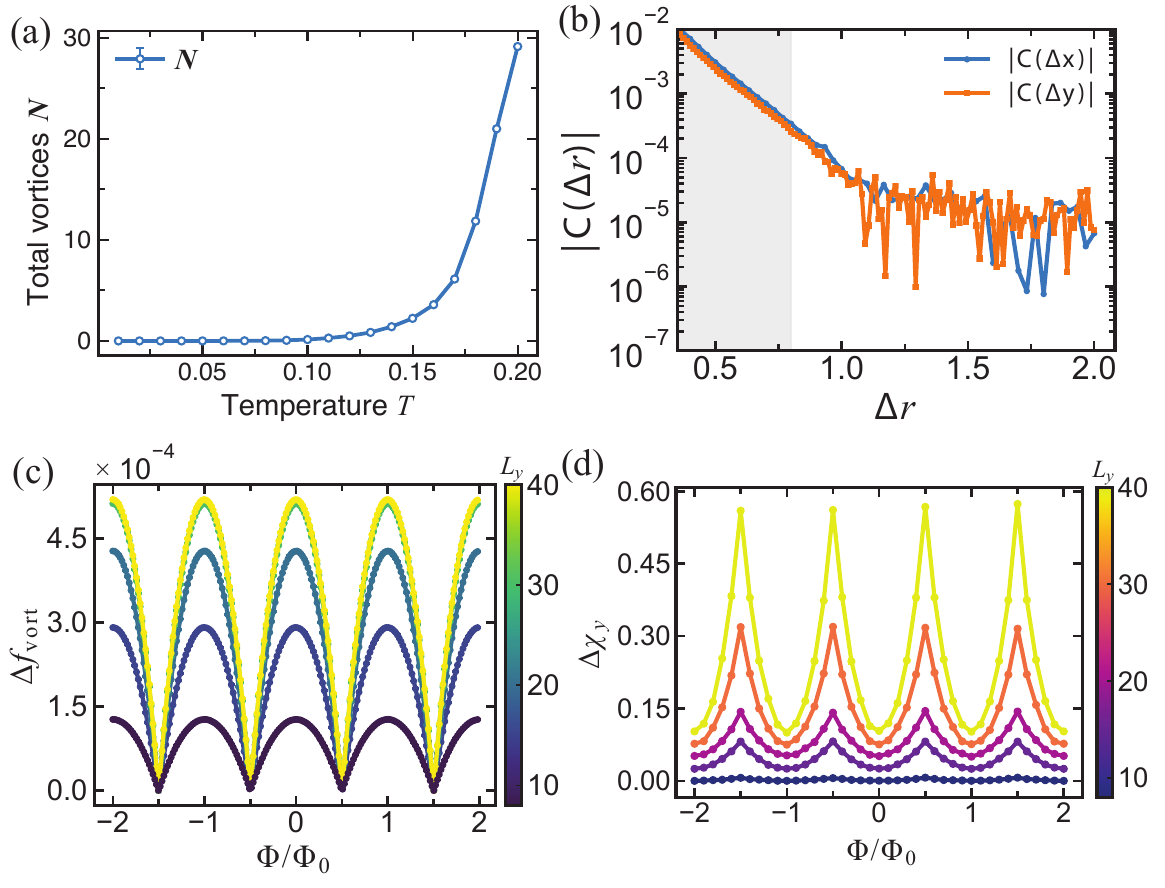}
    \caption{ \addYM{(a) Mean total number of vortices $N$ as a function of temperature $T$. (b) Magnitudes of the  vortex charge-density correlations along the $x$ and $y$ directions, $|C(\Delta x)|$ and $|C(\Delta y)|$, respectively. The correlation function is defined as $C(\mathbf{r},\mathbf{r}')=\langle\rho(\mathbf{r})\rho(\mathbf{r}')\rangle-\langle\rho(\mathbf{r})\rangle\langle\rho(\mathbf{r}')\rangle$, where $\rho(\mathbf{r})$ is the vortex charge density. The shaded region indicates the short-distance fitting range used to extract the correlation lengths. (c) Flux-dependent vortex contribution to the free-energy density, $\Delta f_{\mathrm{vort}}$, as a function of $\Phi/\Phi_0$. (d) Corresponding change in the polarization susceptibility along the $y$ direction, $\Delta\chi_y(\Phi)=\chi_y(\Phi)-\chi_y(0)$, where $\chi_y=\beta\int dydy'(y-L_y/2)(y'-L_y/2)C(y,y')$ and $\beta=1/T$. Here, $C(y,y')$ is obtained from the connected charge-density correlator by summing or averaging over the $x$ direction. Panels (c) and (d) show results for several transverse system sizes $L_y$, as indicated by the color scales.}
}
    \label{Fig_S1}
\end{figure}

\begin{figure}
    \centering
    \includegraphics[width=0.8\linewidth]{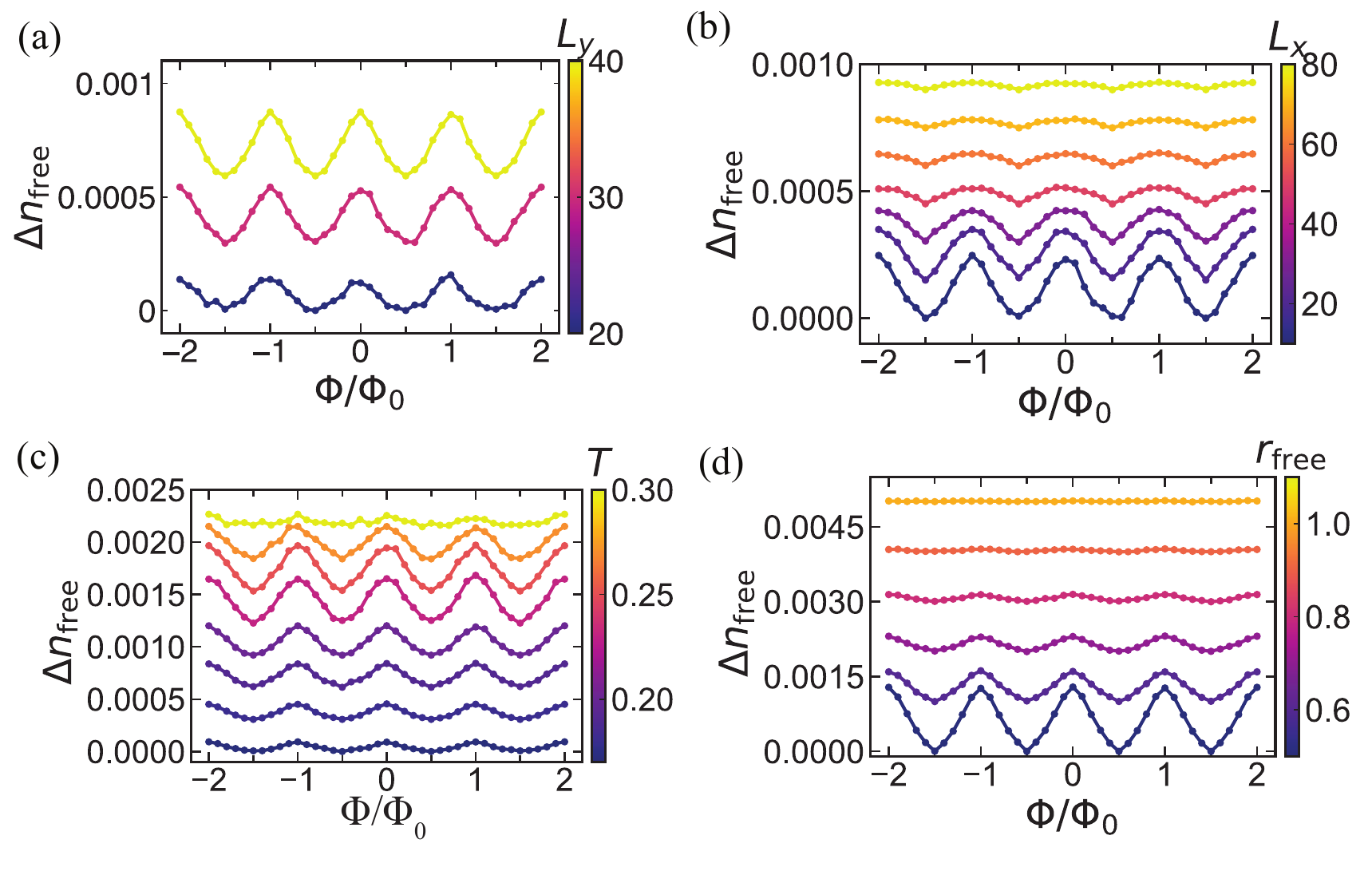}
    \caption{
    {
Flux dependence of the free-vortex density $\Delta n_{\mathrm{free}}$ obtained using the counting method with $r_{\mathrm{free}}$.  The free-vortex density is then obtained by averaging the total number of free vortices over the Monte Carlo configurations and dividing by the system area, $n_{\mathrm{free}}=\langle N_{\mathrm{free}}\rangle/(L_xL_y)$.
Panels (a)--(d) show $\Delta n_{\mathrm{free}}$ as a function of
$\Phi/\Phi_0$ for different values of $L_y$, $L_x$, $T$, and the
free-vortex cutoff distance $r_{\mathrm{free}}$, respectively.
The parameters are (a) $L_x=10$ and $T=0.2$;
(b) $L_y=30$ and $T=0.2$;
(c) $L_x=10$ and $L_y=40$;
and (d) $L_x=10$, $L_y=40$, and $T=0.2$. In particular, 
(d) shows the change in the free vortex density $\Delta n_{\mathrm{free}}$ as a
function of $\Phi/\Phi_0$ for different values of $r_{\mathrm{free}}$.
The amplitude of the periodic oscillations gradually decreases as
$r_{\mathrm{free}}$ increases. However, the positions of the peaks and dips
are insensitive to $r_{\mathrm{free}}$, and the $\pi$-ring behavior remains
robust against variations in $r_{\mathrm{free}}$.
}}
    \label{fig:figs2}
\end{figure}

Superposing the two boundary contributions at $y=0$ and $y=L_y$ gives the total
one--body potential
\begin{equation}
V_{\rm bnd}(y)
= -\lambda\,\Gamma(L_y-y,L_x) + \lambda\,\Gamma(y,L_x).
\end{equation}
Introducing the total boundary charge per line,
$q_{\rm bnd} = \lambda L_x$, one obtains the exact form
\begin{equation}
V_{\rm bnd}(y)
= -\frac{q_{\rm bnd}}{L_x}\Big[\Gamma(L_y-y,L_x) - \Gamma(y,L_x)\Big].
\label{eq:Vbnd_exact}
\end{equation}

\addYM{
In the following, we show that, in a certain limit, the boundary charge effectively behaves as a linear potential along the y direction.  We expand Eq.~\eqref{eq:Phi_exact} for \(L_x\gg a\) using
\[
\arctan\!\frac{L_x}{a} = \frac{\pi}{2} - \frac{a}{L_x} + O\!\big(\tfrac{a^3}{L_x^3}\big),
\qquad
\ln(L_x^2+a^2) = 2\ln L_x + \frac{a^2}{L_x^2} + O\!\big(\tfrac{a^4}{L_x^4}\big).
\]
This yields
\begin{equation}
\Gamma(a,L_x)
= \big(L_x\ln L_x - L_x\big)
+ \frac{\pi}{2}\,a \;-\; \frac{a^2}{2L_x}
+ O\!\Big(\frac{a^4}{L_x^3}\Big).
\label{eq:Phi_asymp}
\end{equation}
Hence, for \(a=y\) and \(a=L_y-y\),
\begin{align}
\Gamma(L_y-y,L_x) - \Gamma(y,L_x)
&= \frac{\pi}{2}(L_y-2y)\;
-\;\frac{(L_y-y)^2 - y^2}{2L_x}\;
+ O\!\Big(\frac{L_y^4}{L_x^3}\Big)\nonumber\\[4pt]
&= \frac{\pi}{2}(L_y-2y)
\;-\;\frac{L_y^2-2L_y y}{2L_x}
\;+\; O\!\Big(\frac{L_y^4}{L_x^3}\Big).
\label{eq:Phi_diff_asymp}
\end{align}
The leading term is \emph{linear in \(y\)} and of order \(O(L_y)\), whereas the next term
is \(O(L_y^2/L_x)\). Therefore,
\begin{equation}
V_{\rm bnd}(y)
\approx -\frac{q_{\rm bnd}}{L_x}\,\frac{\pi}{2}\,(L_y-2y)
\;+\; O\!\Big(\frac{q_{\rm bnd}L^2_y}{L_x^2}\Big)
\label{eq:Vbnd_linear}
\end{equation}
For \(L_x\gg L_y\), the exact potential produced by two \emph{uniform} edge
lines is a \emph{uniform field} (linear in \(y\)), with corrections suppressed by
\(L_y/L_x\).}

{
\subsection{Free energy calculation}

The free energy is formally defined as
\begin{equation}
    F(\phi)=-T\ln Z(\phi),
\end{equation}
where $\phi=\Phi/\Phi_0$ and $Z(\phi)$ is the partition function. In the Monte Carlo simulation, the flux dependence enters through the folded boundary charge
\begin{equation}
    q_{\rm bnd}(\phi)
    =
    \phi-\left\lfloor \phi+\frac{1}{2}\right\rfloor .
\end{equation}
Therefore, the flux-dependent free energy can be written as
\begin{equation}
    F(\phi)=F\!\left(q_{\rm bnd}(\phi)\right).
\end{equation}

The total charge density in the Coulomb-gas representation is decomposed as $\rho_{\rm tot}(\mathbf r)
    =
    \rho_v(\mathbf r)
    +
    \rho_v^{(A)}(\mathbf r)$.
 The Coulomb energy thus separates into three contributions,
\begin{equation}
    E
    =
    E_{vv}
    +
    E_{vA}
    +
    E_{AA},
\end{equation}
Here, $v$ labels the bulk vortex charges, and $A$ labels the boundary vortex charges.

Accordingly, the total flux-dependent free energy is written as
\begin{equation}
    F_{\rm tot}(q_{\rm bnd})
    =
    F_{AA}(q_{\rm bnd})
    +
    F_{\rm vort}(q_{\rm bnd}).
\end{equation}
Here
\begin{equation}
    F_{AA}(q_{\rm bnd})
    =
    E_{AA}(q_{\rm bnd})
\end{equation}
is the bare boundary-charge self-energy, while
\begin{equation}
    F_{\rm vort}(q_{\rm bnd})
    =
    -T
    \ln
    \sum_{\mathcal C}
    \exp\left[
        -\beta
        \left(
            E_{vv}(\mathcal C)
            +
            E_{vA}(\mathcal C;q_{\rm bnd})
        \right)
    \right]
\end{equation}
is the vortex contribution sampled by the Monte Carlo simulation. Thus $F_{\rm vort}$ contains both the bulk--bulk charge interaction and the bulk--boundary charge interaction. The only flux-dependent term not sampled by the Monte Carlo configurations is $F_{AA}$, because it depends only on the imposed boundary charge and not on the vortex positions.

The boundary self-energy can be written as
\begin{equation}
    F_{AA}(q_{\rm bnd})
    =
    \frac{1}{2}K_{AA}q_{\rm bnd}^2,
\end{equation}
with the bare flux stiffness
\begin{equation}
    K_{AA}
    =
    \frac{2}{L_x}
    \left[
        \Gamma(L_y,L_x)
        -
        \Gamma(0,L_x)
    \right].
\end{equation}
The bare superfluid stiffness $\rho_{s}^{0} $ is expected to be proportional to $K_{AA}$ with $\rho_s^0
    =
    \frac{L_x}{4\pi^2 L_y}
    K_{AA}$.  
Since $F_{AA}$ is independent of the vortex configuration, it does not affect the Metropolis acceptance probability at fixed $q_{\rm bnd}$.

The vortex-boundary contribution to the energy can be written as
\begin{equation}
    E_{vA}
    =
    \sum_i q_i V_{\rm bnd}(y_i),
\end{equation}
where $q_i=\pm1$ is the charge of the $i$th vortex. Using Eq.~\eqref{eq:Vbnd_exact}, one obtains
\begin{equation}
    X
    \equiv
    \frac{\partial E_{vA}}{\partial q_{\rm bnd}}
    =
    -\frac{1}{L_x}
    \sum_i q_i
    \left[
        \Gamma(L_y-y_i,L_x)
        -
        \Gamma(y_i,L_x)
    \right].
\end{equation}
This quantity can be measured directly as a thermal average in the Monte Carlo simulation. We have
\begin{equation}
    \frac{\partial F_{\rm vort}}{\partial q_{\rm bnd}}
    =
    \left\langle
    X
    \right\rangle_{q_{\rm bnd}}.
\end{equation}
Although $E_{vv}$ has no explicit dependence on $q_{\rm bnd}$ for a fixed vortex configuration, the equilibrium vortex configurations themselves depend on $q_{\rm bnd}$ through the Boltzmann weight. This indirect dependence is already included in the thermal average
$\langle X\rangle_{q_{\rm bnd}}$, while the explicit derivative entering
$\partial F_{\rm vort}/\partial q_{\rm bnd}$ comes only from $E_{vA}$.
The vortex free-energy density difference is obtained by thermodynamic integration:
\begin{equation}
    \Delta f_{\rm vort}
    =
    f_{\rm vort}(q)-f_{\rm vort}(q_0)
    =
    \frac{1}{L_xL_y}
    \int_{q_0}^{q}
    dq'\,
    \left\langle
    X
    \right\rangle_{q'} .
\end{equation}

The total free-energy density difference is therefore
\begin{equation}
    \Delta f_{\rm tot}(q)
    =
    \frac{
        F_{AA}(q)-F_{AA}(q_0)
    }{L_xL_y}
    +
    \Delta f_{\rm vort}(q).
\end{equation}
Since the absolute value of the free energy is defined only up to an additive constant, we plot either $f(q)-f(q_0)$ or $f(q)-f_{\min}$ as the free-energy oscillation.}

{
Figure~\ref{Fig_S1}(c) shows the vortex contribution to the free energy, including both
the bulk--bulk and bulk--boundary charge interactions. As the flux moves away from an integer value, the free
energy decreases because the vortices rearrange to screen the boundary charges,
producing a dipolar polarization along the $y$ direction. Figure~\ref{Fig_S1}(d) confirms
this behavior through the polarization susceptibility
$\chi_y=\beta\int dy\,dy'\left(y-\frac{L_y}{2}\right)
\left(y'-\frac{L_y}{2}\right)C(y,y')$.
The vortex charge correlation function is
$C(r,r')=\langle\rho(r)\rho(r')\rangle
-\langle\rho(r)\rangle\langle\rho(r')\rangle$,
where $\rho(r)$ is the vortex charge density.}

{
\section{Finite-cylinder Coulomb-gas interaction}

We consider a two-dimensional Coulomb gas on a finite cylinder. The
circumferential direction \(x\) is periodic,
\begin{equation}
    x \equiv x+L_x ,
    \label{eq:x_periodic}
\end{equation}
while the axial direction \(y\) is open,
\begin{equation}
    0<y<L_y .
    \label{eq:y_open}
\end{equation}
A charge or vortex \(i\) has position
\begin{equation}
    \bm r_i=(x_i,y_i),
    \label{eq:position}
\end{equation}
and charge \(q_i=\pm 1\). We define
\begin{equation}
    \Delta x_{ij}=x_i-x_j,
    \qquad
    \Delta y_{ij}=y_i-y_j .
    \label{eq:delta_def}
\end{equation}
For the periodic direction, the minimum-image displacement is
\begin{equation}
    \Delta x_{ij}^{\rm min}
    =
    \Delta x_{ij}
    -
    L_x\,{\rm round}
    \left(
    \frac{\Delta x_{ij}}{L_x}
    \right).
    \label{eq:min_image}
\end{equation}
This definition chooses the shortest representative of the separation along
the circumference, so that
\(\Delta x_{ij}^{\rm min}\in[-L_x/2,L_x/2)\). It avoids spurious large
distances for particles located on opposite sides of the coordinate boundary.

The local two-dimensional Coulomb interaction is
\begin{equation}
    V_{ij}^{\rm log}
    =
    -q_iq_j
    \log
    \left[
    \frac{
    \sqrt{
    \left(\Delta x_{ij}^{\rm min}\right)^2
    +
    \left(\Delta y_{ij}\right)^2
    }
    }{r_c}
    \right],
    \label{eq:bare_log}
\end{equation}
where \(r_c\) is the short-distance core cutoff. This is the simplest
minimum-image logarithmic potential.

{\bf Periodic-\texorpdfstring{\(x\)}{x} correction to  the logarithmic potential.} The exact Green's function for a system periodic in \(x\) and infinite in
\(y\) can be written in terms of the periodic-cylinder distance
\begin{equation}
    D_x(\Delta x,\Delta y)
    =
    \frac{L_x}{\pi}
    \sqrt{
    \sin^2
    \left(
    \frac{\pi \Delta x}{L_x}
    \right)
    +
    \sinh^2
    \left(
    \frac{\pi \Delta y}{L_x}
    \right)
    } .
    \label{eq:Dx}
\end{equation}
The corresponding pair interaction is
\begin{equation}
    V_{ij}^{x{\rm -per}}
    =
    -q_iq_j
    \log
    \left[
    \frac{
    D_x(\Delta x_{ij},\Delta y_{ij})
    }{r_c}
    \right].
    \label{eq:x_periodic_potential}
\end{equation}
The \(\sinh\) term appears because \(x\) is periodic while \(y\) is not.
It is not an image-charge contribution from an open boundary; rather, it
comes from summing the periodic copies in the circumferential direction.

Equivalently, Eq.~\eqref{eq:x_periodic_potential} can be written as the
bare logarithmic interaction plus a periodic-\(x\) correction,
\begin{equation}
    V_{ij}^{x{\rm -per}}
    =
    V_{ij}^{\rm log}
    +
    V_{ij}^{x{\rm -corr}},
    \label{eq:xper_decomp}
\end{equation}
where
\begin{align}
    V_{ij}^{x{\rm -corr}}
    =
    -q_iq_j
    \bigg\{
    &
    \log
    \left[
    \frac{
    D_x(\Delta x_{ij},\Delta y_{ij})
    }{r_c}
    \right]
    \nonumber\\
    &
    -
    \log
    \left[
    \frac{
    \sqrt{
    \left(\Delta x_{ij}^{\rm min}\right)^2
    +
    \left(\Delta y_{ij}\right)^2
    }
    }{r_c}
    \right]
    \bigg\}.
    \label{eq:x_corr}
\end{align}

{\bf Open-\texorpdfstring{\(y\)}{y} image-charge contribution.} The open ends at \(y=0\) and \(y=L_y\) can be treated by image charges.
The sign of the image charge depends on the physical boundary condition. For
the opposite-sign image construction, the image of a charge \(q_j\) at
\((x_j,y_j)\) with respect to the lower boundary is located at
\begin{equation}
    (x_j,-y_j),
    \label{eq:bottom_image_position}
\end{equation}
and has charge \(-q_j\). The corresponding lower-boundary image contribution is
\begin{equation}
    V_{ij}^{\rm image,bottom}
    =
    +q_iq_j
    \log
    \left[
    \frac{
    D_x(\Delta x_{ij},y_i+y_j)
    }{r_c}
    \right].
    \label{eq:image_bottom}
\end{equation}
Similarly, the first image with respect to the upper boundary at \(y=L_y\) is
located at
\begin{equation}
    (x_j,2L_y-y_j),
    \label{eq:top_image_position}
\end{equation}
giving
\begin{equation}
    V_{ij}^{\rm image,top}
    =
    +q_iq_j
    \log
    \left[
    \frac{
    D_x(\Delta x_{ij},y_i+y_j-2L_y)
    }{r_c}
    \right].
    \label{eq:image_top}
\end{equation}

Thus, in a first-image approximation, the finite-cylinder interaction is
\begin{align}
    V_{ij}^{\rm finite}
    \approx
    &
    -q_iq_j
    \log
    \left[
    \frac{
    D_x(\Delta x_{ij},y_i-y_j)
    }{r_c}
    \right]
    \nonumber\\
    &
    +q_iq_j
    \log
    \left[
    \frac{
    D_x(\Delta x_{ij},y_i+y_j)
    }{r_c}
    \right]
    \nonumber\\
    &
    +q_iq_j
    \log
    \left[
    \frac{
    D_x(\Delta x_{ij},y_i+y_j-2L_y)
    }{r_c}
    \right].
    \label{eq:first_image}
\end{align}

A more complete two-boundary image construction includes repeated reflections.
For opposite-sign images, one convenient finite-\(M\) expression is
\begin{align}
    V_{ij}^{\rm finite}
    =
    -q_iq_j
    \sum_{m=-M}^{M}
    \bigg[
    &
    \log
    \left(
    \frac{
    D_x(\Delta x_{ij},y_i-y_j+2mL_y)
    }{r_c}
    \right)
    \nonumber\\
    &
    -
    \log
    \left(
    \frac{
    D_x(\Delta x_{ij},y_i+y_j+2mL_y)
    }{r_c}
    \right)
    \bigg].
    \label{eq:image_series}
\end{align}
The limit \(M\rightarrow\infty\) gives the full image series for this boundary
condition. In numerical calculations, \(M\) can be truncated once the result is
converged.

In addition to the pairwise image interaction, each charge interacts with its
own images. This produces a one-body self-image potential. For the
opposite-sign image construction, a finite-\(M\) expression is
\begin{align}
    V_i^{\rm self-image}
    =
    -\frac{1}{2}q_i^2
    \sum_{m=-M}^{M}
    \bigg[
    &
    {}
    \log
    \left(
    \frac{
    D_x(0,2mL_y)
    }{r_c}
    \right)
    \nonumber\\
    &
    -
    \log
    \left(
    \frac{
    D_x(0,2y_i+2mL_y)
    }{r_c}
    \right)
    \bigg].
    \label{eq:self_image}
\end{align}
The prime indicates that the singular real self term with \(m=0\) is omitted.

The total finite-cylinder energy can therefore be written as
\begin{equation}
    E
    =
    \sum_{i<j} V_{ij}^{\rm finite}
    +
    \sum_i V_i^{\rm self-image}
    +
    E_{\rm ext},
    \label{eq:total_energy}
\end{equation}
where \(E_{\rm ext}\) denotes any additional external or boundary-background
potential.

{\bf Relation to the logarithmic interaction simulations.}
In the previous simulations, we used an approximate logarithmic pair interaction,
\begin{equation}
    V_{ij}^{\rm sim}
    =
    -q_iq_j
    \log
    \left[
    \frac{
    \sqrt{
    \left(\Delta x_{ij}^{\rm min}\right)^2
    +
    \left(\Delta y_{ij}\right)^2
    }
    }{r_c}
    \right].
    \label{eq:sim_potential}
\end{equation}
Here \(\Delta x_{ij}^{\rm min}\) imposes the cylindrical topology at the
level of the minimum-image convention along the periodic circumferential
direction. Therefore, Eq.~\eqref{eq:sim_potential} captures the short-distance
logarithmic Coulomb interaction and treats particles close across the
periodic boundary as nearby.

However, Eq.~\eqref{eq:sim_potential} should be viewed as a logarithmic
minimum-image approximation rather than the full Green's function of a finite
cylinder. The full finite-cylinder interaction can be organized as
\begin{equation}
    V_{ij}^{\rm finite}
    =
    V_{ij}^{\rm log}
    +
    V_{ij}^{x{\rm -corr}}
    +
    V_{ij}^{y{\rm -image}},
    \label{eq:finite_decomposition}
\end{equation}
where \(V_{ij}^{\rm log}\) is the local logarithmic interaction,
\(V_{ij}^{x{\rm -corr}}\) corrects the minimum-image form to the
periodic-\(x\) Green's-function kernel, and \(V_{ij}^{y{\rm -image}}\)
accounts for image charges associated with the open boundaries at
\(y=0\) and \(y=L_y\).

The last two terms, \(V_{ij}^{x{\rm -corr}}\) and
\(V_{ij}^{y{\rm -image}}\), are finite-geometry corrections. They do not
modify the local logarithmic interaction at short distances, since the
periodic-\(x\) kernel reduces to the ordinary distance in this case, and the image terms are subleading for
vortices far from the open boundaries. Therefore, these corrections are not
expected to change the local Coulomb-gas physics. However, they may affect
quantitative finite-size effects, long-range interactions, and vortex
configurations close to the open ends of the cylinder.}

 \end{document}